\documentclass[aps,pre,twocolumn,superscriptaddress,showkeys]{revtex4-2}

\usepackage[T1]{fontenc}
\usepackage{amsmath}
\usepackage{amssymb}
\usepackage{amsthm}
\usepackage{amsfonts}
\usepackage{graphicx}
\usepackage{blindtext}
\usepackage{xcolor}
\usepackage{verbatim}
\usepackage{dcolumn}
\usepackage{bm}
\usepackage{float}
\usepackage{color}
\definecolor{rojo}{rgb}{1,0,0}
\definecolor{verde}{rgb}{0,0.8,0.2}
\definecolor{azul}{rgb}{0,0,1}
\definecolor{rosa}{cmyk}{0,1,0,0}
\usepackage{appendix}
\usepackage{epsfig}
\usepackage{hyperref}
\hypersetup{colorlinks=true, citecolor=blue}
\usepackage{soul}
\hypersetup{
     colorlinks=true,
     linkcolor=blue,
     filecolor=blue,
     citecolor = blue,      
     urlcolor=blue,
     }

\begin{document}
\preprint{APS/123-QED}

\title{Bulk Modulus along Jamming Transition Lines
of Bidisperse Granular Packings}

\author{Juan C. Petit}
\author{Nishant Kumar}%
\affiliation{Institut f\"{u}r Materialphysik im Weltraum, Deutsches Zentrum f\"{u}r Luft- und Raumfahrt (DLR), 51170 K\"{o}ln, Germany }%

\author{Stefan Luding}%
\affiliation{Multi-Scale Mechanics (MSM), TFE-ET, MESA+, University of Twente, Enschede, Netherlands }%

\author{Matthias Sperl}%
\affiliation{Institut f\"{u}r Materialphysik im Weltraum, Deutsches Zentrum f\"{u}r Luft- und Raumfahrt (DLR), 51170 K\"{o}ln, Germany }%
\affiliation{Institut f\"{u}r Theoretische Physik, Universit\"{a}t zu K\"{o}ln, 50937 K\"{o}ln, Germany}%

\begin{abstract}
We present 3D DEM simulations of bidisperse granular packings to investigate their jamming densities, $\phi_J$, 
and dimensionless bulk moduli, $K$, as function of the size ratio, $\delta$, and the concentration of small particles, $X_{\mathrm S}$. 
We determine the partial and total bulk moduli for packings near their jamming densities, including a second transition
that occurs for sufficiently small $\delta$ and $X_{\mathrm{S}}$ when the system is compressed
beyond its first jamming transition. While the first transition is sharp, exclusively with large-large contacts, 
the second is rather smooth, carried by small-large interactions at
densities much higher than the mono disperse random packing baseline,
$\phi_J^{\rm{mono}} \approx 0.64$. When only non-rattlers are considered, all the effective transition 
densities are reduced, and the density of the second transition emerges rather close to the 
reduced base-line, $\tilde{\phi}_J^{\rm mono} \approx 0.61$, due to its smooth nature.
At size ratios $\delta \le 0.22$ a concentration $X^{*}_{\mathrm S}$ divides the diagram:
either with most small particles non-jammed, or jammed jointly with large ones. 
For $X_{\mathrm S} < X^{*}_{\mathrm S}$, the modulus $K$ displays different behaviors at first and second 
jamming transitions. Along the second transition, $K$ rises relative to the values found at the first 
transition, however, is still small compared to $K$ at $X^{*}_{\mathrm S}$. 
Explicitly, for our smallest $\delta = 0.15$, the discontinuous jump in $K$ as a function of 
$X_{\mathrm S}$ is obtained at $X^{*}_{\mathrm S} \approx 0.21$ -- and coincides with the maximum 
jamming density where both particle species mix most efficiently. Our new results will 
allow tuning or switching the bulk modulus $K$ or other properties, such as the wave speed, by choosing specific sizes 
and concentrations based on a better understanding of whether small particles contribute to the jammed 
structure or not, and how the micromechanical structure behaves at either transition. 
\end{abstract}

\maketitle

\section{INTRODUCTION}

When a collection of non-touching spheres is externally compressed, there is a 
critical packing fraction, at which the contacts between spheres percolate 
across the whole system \cite{ziff2017percolation, aharonov1999rigidity, pathak2017force, 
majmudar2007jamming, kumar2016memory}. 
At this state, the granular packing develops a 
mechanically stable 
structure that can reversibly withstand further external deformation
-- at least for small enough (infinitesimal) strains \cite{luding2021jamming}. 
Such a state is known as the jammed 
state and is characterized by a jamming density $\phi_J$. 
For stiff, rigid particles, one can talk about ``rigidity'', 
whereas, for soft particles, the structure and bulk properties
of granular packings have been quantified by mechanical properties such as the dimensionless bulk 
modulus, $K$. Close enough to jamming, for small enough confining stress
and thus small enough contact deformations, the difference between soft and stiff should diminish,
while only soft particles allow exploring the jammed state by increasing the deformations
\cite{luding2021jamming}.
Previous works have shown that $\phi_{J}$ and $K$ depend 
not only on the size ratio, $\delta$, and on the concentration of small 
particles $X_{\mathrm S}$ \cite{prasad2017subjamming,hopkins2011phase,frankrichter2014disordered, 
biazzo2009theory, pillitteri2019jamming, kumar2016memory, kumar2016tuning, hara2021phase, koeze2016mapping, suo2022unexplored}, 
but also on the preparation procedure \cite{kumar2016memory, luding2021jamming, luding2021does}. 
For example, this allows tuning the packing density and effective bulk modulus of frictionless 
particles to the highest values by choosing different combinations of ($\delta$, $X_{\mathrm S}$)
for bidisperse packings \cite{kumar2016tuning, petit2020additional}. The multitude of 
generally possible size distributions (see Ref.~\cite{ogarko2013prediction} and references therein) 
is not considered here.

In a recent paper \cite{petit2020additional}, we have
explored in detail the impact of a wide range of $\delta$ and $X_{\mathrm S}$ on jamming.
Upon compression, bidisperse packings experience an additional transition at low 
$\delta$ and low $X_{\mathrm S}$. Two transitions  
arise: one driven by predominantly large particles obtained at low densities and the other 
by jamming small particles jointly with large ones at higher densities upon futher compression. 
This means that the second transition is the source of an additional stiffness of the 
jammed packing, rather smooth compared to the sharp first transition.
This additional transition has opened a new window of research since  
packing structures obtained along this additional line might lead to 
different mechanical properties. Our aim in this work is to study the mechanical 
properties of bidisperse systems along the jamming lines, especially along the 
additional transition line found in Ref.~\cite{petit2020additional}. 
Here, we focus on the bulk modulus to quantify both the rigidity of jammed 
bidisperse packings and also the second transition at higher densities, 
as it is accessible from isotropic compression. 
This means that the shear modulus cannot be calculated from this protocol,
at least not without additional probing, e.g., as done in Ref.~\cite{kumar2014macroscopic}. 
We also analyze the relevance of rattlers in jammed bidisperse 
granular packings and the number of contacts along the jamming lines. In general, this 
work attempts to answer the following questions: \\
$\bullet$ How does the bulk modulus $K$ of a 
jammed bidisperse packing change as a function of $\delta$ and $X_{\mathrm S}$? \\
$\bullet$ Which values of ($\delta$, $X_{\mathrm S}$) provide the highest $K$? \\
$\bullet$ What are the relations between macro-variables like $K$ with the 
micro-structural features (packing fraction, contact number, 
fabric, force networks)?\\

This paper is organized as follows. In Sec.~\ref{SecII}, we discuss the technique 
and procedure of the simulation. We define the concentration of small particles, $X_{\mathrm S}$, 
and discuss how the number of particles in each bidisperse mixture changes with $X_{\mathrm S}$. 
In Sec.~\ref{SecIII}, we explain the jamming diagram and discuss the nature of the second 
transition line found at low $\delta$ and low $X_{\mathrm S}$. Sec.~\ref{SecIV} shows 
the relevance of rattlers on the jamming density and mean contact number along the jamming lines. 
Sec.~\ref{SecV} presents the dependence of the effective bulk modulus at and above jamming on $\delta$ and $X_{\mathrm S}$. 
We conclude with a summary and further discussions.

\section{Technique and procedure of the simulation}
\label{SecII}

\begin{figure}[t]
	\centering \includegraphics[scale=0.33]{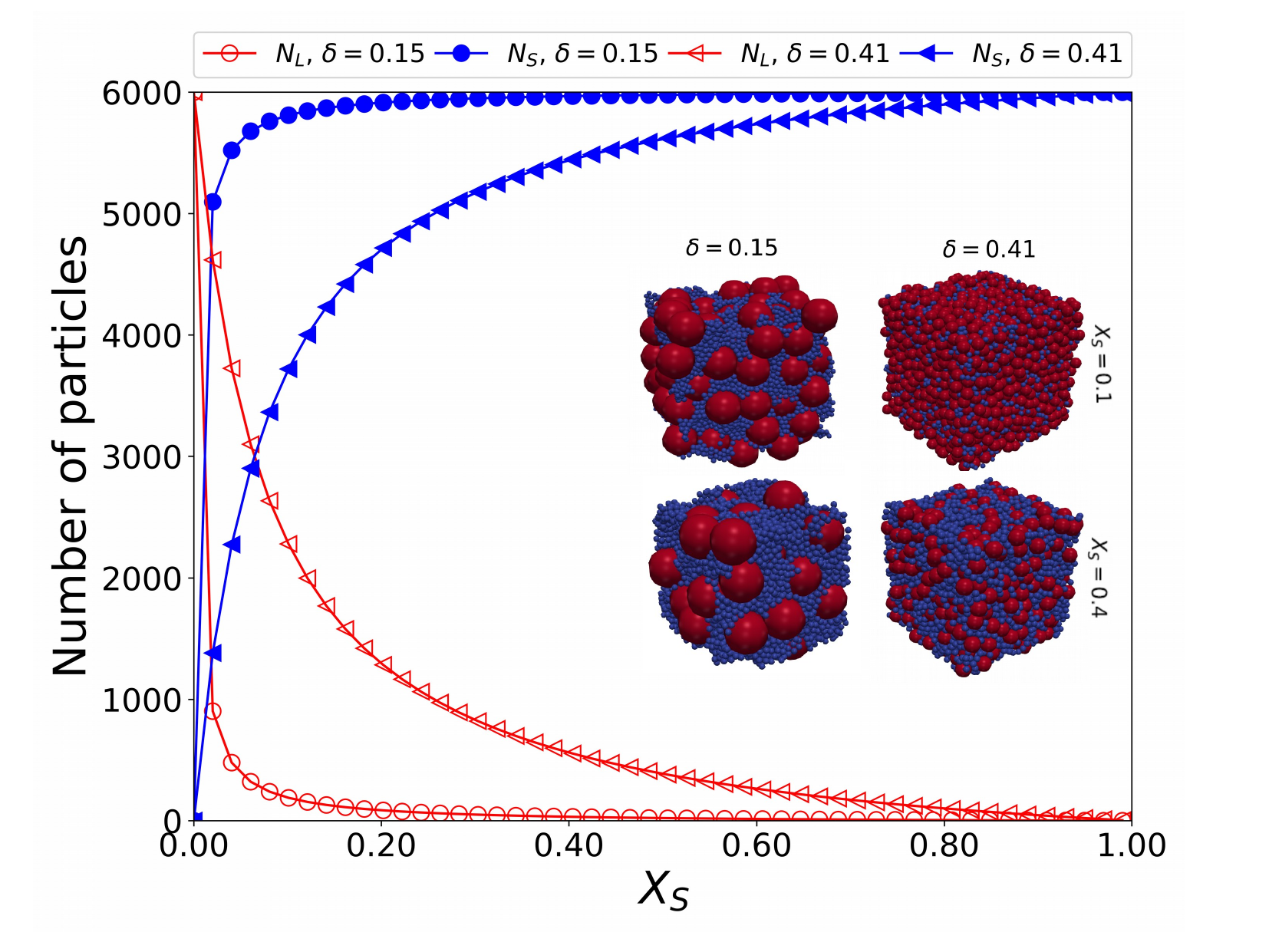}
	\caption{Number of large, $N_{\mathrm L}$,  and small, $N_{\mathrm S}$, particles as a function of 
	$X_{\mathrm S}$ for two typical $\delta$. The total number of particles is set to $N = 6000$. The inset exhibits
	four combinations of $(\delta, X_{\mathrm S})$ corresponding to jammed packings at the maximum compression, 
	$\phi_{\rm{max}}=0.90$. }
	\label{packings}
\end{figure}

We use the open source code {\tt www.MercuryDPM.org} 
to perform 3D Discrete Element Method (DEM) simulations 
to study the macroscopic properties of soft-sphere jammed packings \cite{cundall1979discrete, 
luding2005shear, kruyt2010micromechanical, petit2017contact, petit2018reduction}. 
Newton's equations for all particles are numerically 
solved to analyze their motion in time. $N = N_\mathrm{L}+N_\mathrm{S} = 6000$ 
particles are used to set up bidisperse packings, 
with a number of large, 
$N_{\mathrm L}$, and small, $N_{\mathrm S}$, particles,
with dimensionless radius, $r_{\mathrm L} = 1/2$,
kept constant, and $r_{\mathrm S}$ varied with the size ratio,
$\delta = r_{\mathrm S}/r_{\mathrm L} \in [0.15, 1]$.

In general, we use dimensionless quantities, $\Gamma$,
using the transformation $\Gamma' = \Gamma \Omega'_{u}$, where the prime represents the 
variables with units, variables without prime are dimensionless, and $\Omega'_{u}$ is the 
scale variable carrying the units, see Ref.~\cite{luding2021jamming}.
For example, the transformation of the large particle radius is
$r'_{\mathrm{L}} = r_{\mathrm{L}}x'_{u}$, where we choose the unit of length,
$x'_{u} = 2r'_{\mathrm{L}} = 3$, \footnote{The units of the values used in the computer
are non-specified, but could be read as $x'_u=3$ [mm] for length, 
$\rho'_u=2000$ [kg/m$^3$] for density, $\kappa'_u=10^5$ [kg/s$^2$] 
for stiffness, $\gamma' = 1000$ [kg/s] for viscosity
and therefore, $t_u’ \approx 2.32 \times 10^{-5}$ [s] for time. }
as the large particle diameter, 
i.e., the dimensionless radius of large particles is $r_{\mathrm{L}} = 1/2$.
Similarly, the small particle diameter,
$2r_{\mathrm{S}} = r'_{\mathrm{S}}/r'_{\mathrm{L}}=:\delta$, 
defines the size ratio. 

The same material density is used for large and small particles, 
which makes it convenient to choose the unit of density as:
$\rho'_{u} = \rho'_{p} = 2000$, i.e.,
the dimensionless density is: $\rho_{p} = 1$.
Consequently, the unit of mass is: 
$m'_{u} =  8\rho'_{p} r_{\mathrm{L}}^{\prime 3}$, 
so that the dimensionless masses of 
large and small particles are:
$m_{\mathrm{L}} = \frac{\pi}{6} $ and 
$m_{\mathrm{S}} = \frac{\pi}{6} \delta^{3} = m_{\mathrm{L}}\delta^{3}$, respectively. 

As last free unit, we choose the time: 
$t'_{u} = (m'_{u}/\kappa'_{u})^{1/2}$, 
proportional to the collision time scale $t'_c$ [see subsection \ref{sec:contactmodel} and
Appendix~\ref{AppA}],
with stiffness $\kappa'_{u} = \kappa'_{n} = 10^{5}$, 
so that:  $\kappa_{n} = 1$.
The unit of time follows from the previous $t'_{u} = (8 \rho'_{p}/\kappa'_{n})^{1/2} r_{\mathrm{L}}^{\prime 3/2} \approx 0.735$.
The viscous damping used is $\gamma'_{n} = 1000$, resulting in a dimensionless 
$\gamma_{n} = \gamma'_n /(m'_u/t'_u)
= \gamma'_{n}/(8\rho'_{p}\kappa'_{n} r_{\mathrm{L}}^{\prime 3} )^{1/2}  \approx 0.0136$. 
From our non-dimensionalization, we obtain 
the dimensionless pressure as: 
$P = P' (x'_u/\kappa'_u ) 
   = 3 \times 10^{-5} P'$.
Note that the identity: 
$P = 2 r_{\mathrm{L}} P / \kappa_n = 2 r'_{\mathrm{L}} P' / \kappa'_n$
is a consequence of our choice of units, but not true in general for other choices.

\subsection{Bidisperse systems}

A bidisperse packing is characterized by its 
size ratio, $\delta$, the volume concentration of small particles, 
$X_{\mathrm S} = N_{\mathrm S} \delta^{3} / (N_{\mathrm L} + N_{\mathrm S} \delta^{3})$,
and its density, $\phi_J$. The former
can be controlled, but the latter depends, 
e.g., on the preparation procedure.
Fig.~\ref{packings} shows the variation of the number of large and small particles as 
a function of $ X_{\mathrm S}$ for two typical values of $\delta$. For a particular 
$\delta$, keeping $N$ constant, the number of small particles increases rapidly, while the 
large one decrease, as $X_{\mathrm S} \to 1$. The intersection point, which 
represents packings formed by the same amount of small and large particles, $N_{\mathrm L} = 
N_{\mathrm S} = N/2$, is shifted to lower $X_{\mathrm S}$ as $\delta$ 
decreases, representing the 50:50 bidisperse particle mixture studied previously using $\delta = 0.71$ 
\cite{ohern2003jamming, majmudar2007jamming}. However, 
we will argue in Sec.~\ref{SecV} that using the 50:50 mixture does not provide the 
densest packing and highest bulk modulus, instead, one has to choose a proper combination 
of ($\delta$, $X_{\mathrm S}$). For the case of 
$\delta = 0.15$ the intersection point is as low as $X_{\mathrm S} = 0.01$. 
Below the intersection point shown in Fig.~\ref{packings}, the packing is formed by small 
particles in a sea of large ones (not shown, see \cite{kumar2016tuning}), while 
far above, the few large particles
are embedded in a sea of small ones, 
see the insets for $X_{\mathrm S} = 0.4$.

\subsection{Contact Model}

\label{sec:contactmodel}
We use the linear normal contact force model given as 
${\it \bf f}^n_{ij}=f^n_{ij} \hat{\bf n} = ( \kappa_{n} \alpha_c + \gamma_{n} \dot\alpha_c) \hat{\bf n}$ 
\cite{cundall1979discrete, luding2008cohesive, kumar2016tuning}, with the 
contact overlap $\alpha_{c} = (r_{i}+r_{j}) - a_{ij}$, where $r_{i}$, $r_{j}$ and $a_{ij}$ 
are the radii and relative separation between the centers of particles $i$ and $j$, respectively. 
$\dot\alpha_c$ is the 
relative velocity in the normal 
direction $\hat{\bf n}$. 
An artificial background dissipation force, 
${\it \bf f}_b=-\gamma_{b} {\bf v}_i$, 
with $\gamma_{b} = \gamma_{n}$, proportional to 
the velocity ${\bf v}_i$ of particle $i$ is added, 
resembling the damping due to a background medium. 

From the contact model, 
one can compute [see Appendix~\ref{AppA}]
the contact duration, $t_{c}$, and the restitution 
coefficient, $e$, both depending on $\delta$. 
For example, the fastest response $t_{c}$ corresponds 
to the interaction between the smallest particles 
($\delta = 0.15$) as $t^{\mathrm{SS}}_{c} = t^{'\mathrm{SS}}_{c}/t'_{u} 
= 0.07 (\kappa'_{n}/8\rho'_{p} r_{\mathrm{L}}^{\prime 3})^{1/2} $
$\approx 0.09$, together with the strongest dissipation, 
$e_{\mathrm{SS}} = 0.477$, see Appendix \ref{AppA} for its definition 
and Fig.~\ref{app_fig1} for its variation with $\delta$.
The integration time step is chosen to be 50 times smaller 
than this shortest time scale $t^{\mathrm{SS}}_{c}$. 
If even smaller size ratios $\delta$ would be used, 
computation time would increase since finer time steps are needed to resolve 
the collisions among small particles. We expect similar results for $\delta < 0.15$,
but restrict ourselves to a minimal $\delta = 0.15$ for this study. 
The contact and background dissipation terms are used to 
damp kinetic energy and reduce computational time during relaxation. 

We restrict ourselves to the linear contact model without any friction
between particles \citep{imole2013hydrostatic, kumar2014macroscopic}. 
Thus, we exclude all the non-linearities 
present in the system due to contact models and focus on the effect of the size ratio and 
concentration of small particles on jamming.

\begin{figure}[t]
  \centering
    \includegraphics[scale=0.35]{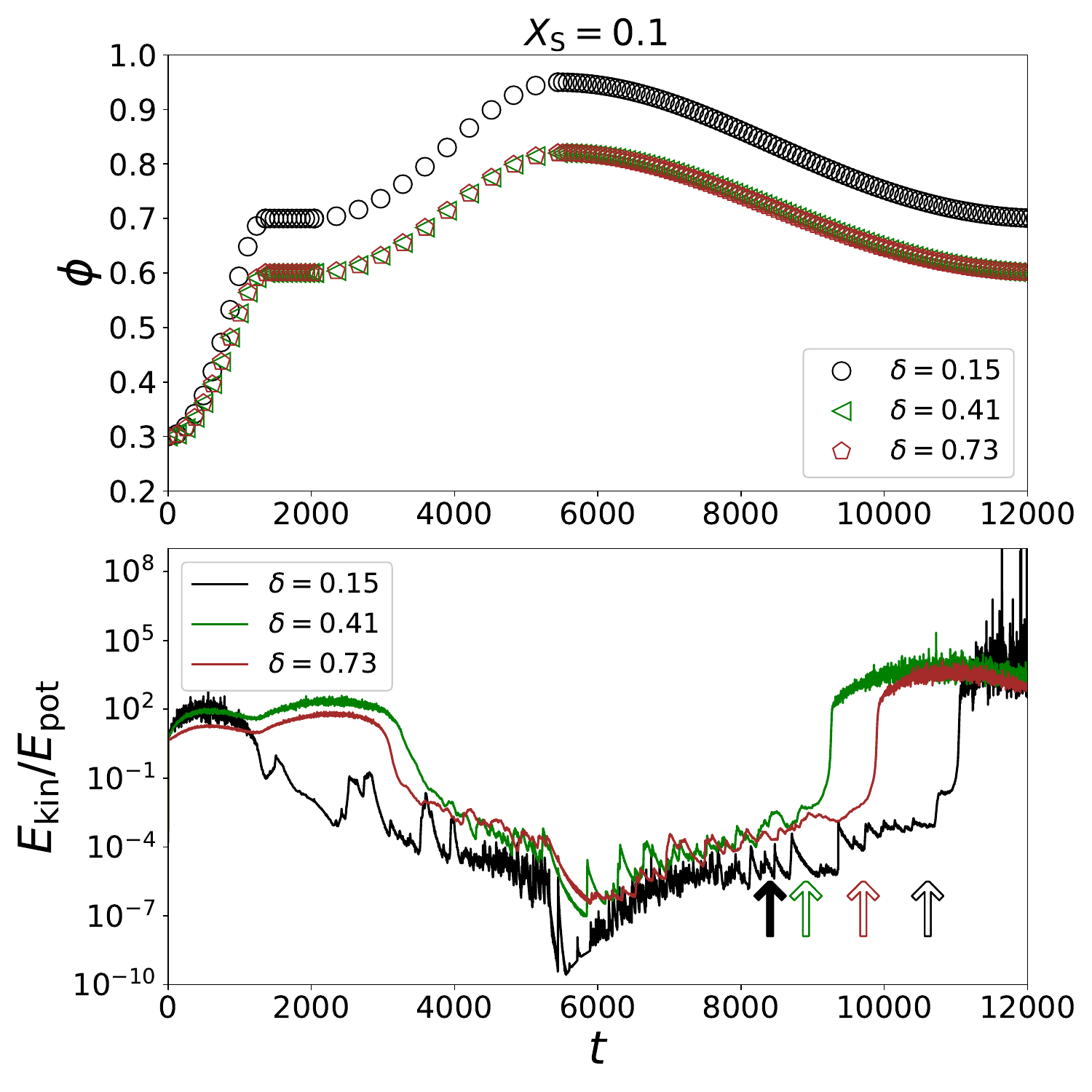}

\caption{
(Top) Illustration of the evolution of packing fraction, $\phi$,
as a function of the dimensionless time, $t$.
The protocol allows accessing both jamming and unjamming densities of the system. 
Note that for $\delta = 0.41,\,0.73$,  the initial target of $\phi_0 = 0.60$ and 
$\phi_{\mathrm{max}} = 0.82$ are chosen, 
whereas for $\delta = 0.15$,  $\phi_0 = 0.70$ and $\phi_{\mathrm{max}} = 0.95$. 
(Bottom) Energy ratio plotted against $t$ at $X_{\mathrm S} = 0.1$ for different $\delta$.
The empty arrows represent where the first jamming transition 
is identified along the decompression branch, while the solid one corresponds to the second transition.  
The increase in energy ratio during the relaxation phase is due to $E_{\rm pot}$ decaying faster 
than $E_{\rm kin}$, for the larger values of $\delta$ that have comparatively weak dissipation.}
\label{protocol}
\end{figure}

\subsection{Testing protocol}

Each bidisperse packing characterized by the parameters ($\delta$, $X_{\mathrm S}$) is created and 
further compressed using a unique, well defined protocol \citep{goncu2010constitutive}, see Fig.~\ref{protocol} (Top). 
The initial configuration is such that spherical particles with dimensionless radii $r_{\mathrm L}$ and 
$r_{\mathrm S}$ are uniform, and randomly placed in a 3D box without gravity, 
with an initial packing fraction of $\phi_{\mathrm{ini}} = 0.3$ 
and uniform random velocities. The possible artificial large overlaps lead to a peak in kinetic energy at 
the beginning, which is quickly damped due to background and collisional dissipation. Such low density 
systems with high kinetic energy only help to quickly randomize the particles. Changes in the initial spatial and velocity distributions should not change the jamming density.
The granular gas is then isotropically compressed, to approach a direction 
independent initial configuration with target packing fraction, $\phi_0 $, which depends on 
$\delta$ and $X_{\mathrm S}$, being slightly below the jamming density, i.e., the transition from fluid-like to solid-like behavior \citep{majmudar2007jamming,ohern2002random,makse2000packing,van2010jamming}.
After this initial preparation, randomization, compression phase, where kinetic energy can 
be considerable to take care that the particle positions are randomized, first, a relaxation process of the system is allowed at constant $\phi=\phi_0$.

The smooth isotropic compression (decompression) up to $\phi = \phi_{\mathrm{max}}$ 
(back to $\phi = \phi_{\mathrm{0}}$) is realized by a simultaneous inward (outward) movement of all 
periodic boundaries of the system. 
For example, the vertical wall position, or height, is given by 
\begin{equation}
z(t) = z_{f} + \frac{z_{0}-z_{f}}{2}(1 + \mathrm{cos}(2\pi\,(t-t_0)/T)),
\label{ecu0} 
\end{equation} 
\noindent 
with strain $\varepsilon_{zz}(t) = 1 - z(t)/z_{0}$, where $z_{0}$ and $z_{f}$ are 
the initial and extreme vertical size of the box at zero and maximum strain, respectively. 
$T$ is the dimensionless total time of simulation defined by $T = T'/t'_{u} \approx 12000$, 
with $T'= 9000$. Note that the simulation procedure observed in 
Fig.~\ref{protocol} is driven by the target packing fraction chose at different stages. 
During the loading/unloading periods the deformation rate is of the order  
$\dot\varepsilon_v= \dot\varepsilon_{xx}+\dot\varepsilon_{yy}+ \dot\varepsilon_{zz}
\propto (1-z_f/z_0)\,\mathrm{sin}(2\pi\,(t-t_0)/T)/T$.

The time-offset, $t_0$, is chosen such that all wall motions by
the co-sinusoidal function allows for a smooth start-up and finish of the motion so that shocks and inertia effects are avoided.
Other testing methods could be used 
\cite{donev2006binary, o2003jamming, chaudhuri2010jamming}, 
however, they should have no different outcome since the deformation is very slow, a regime for which the equivalence between 
our procedure and an energy minimization approach was shown in Ref.~\cite{krijgsman2016simulating}.

Given sufficient energy is dissipated by either background medium or collisions, this allows to determine a consistent jamming density at quite low strain rates, see Ref.~\citep{ogarko2012equation}. To quantify the rate of deformation, it is convenient to define the dimensionless inertial number, $I_v=\dot\varepsilon_v d/\sqrt{P/\rho}$, analogous to the shear inertial number \cite{luding2021jamming}. In the existence of background damping, an additional viscous number, $I_{\gamma}$, together with $I_v$, characterizes the rheology of the system, see Refs.~\cite{boyer2011unifying,vo2020additive}. Note that rheology is not investigated in this study, the contact and
background dissipation are used to remove the energy before the resulting static packings are analyzed.
A brief discussion of the role of the background damping in the dynamics of the systems is given in 
Appendix~\ref{AppA}.

\begin{figure}[t]
    
    \centering \includegraphics[scale=0.3]{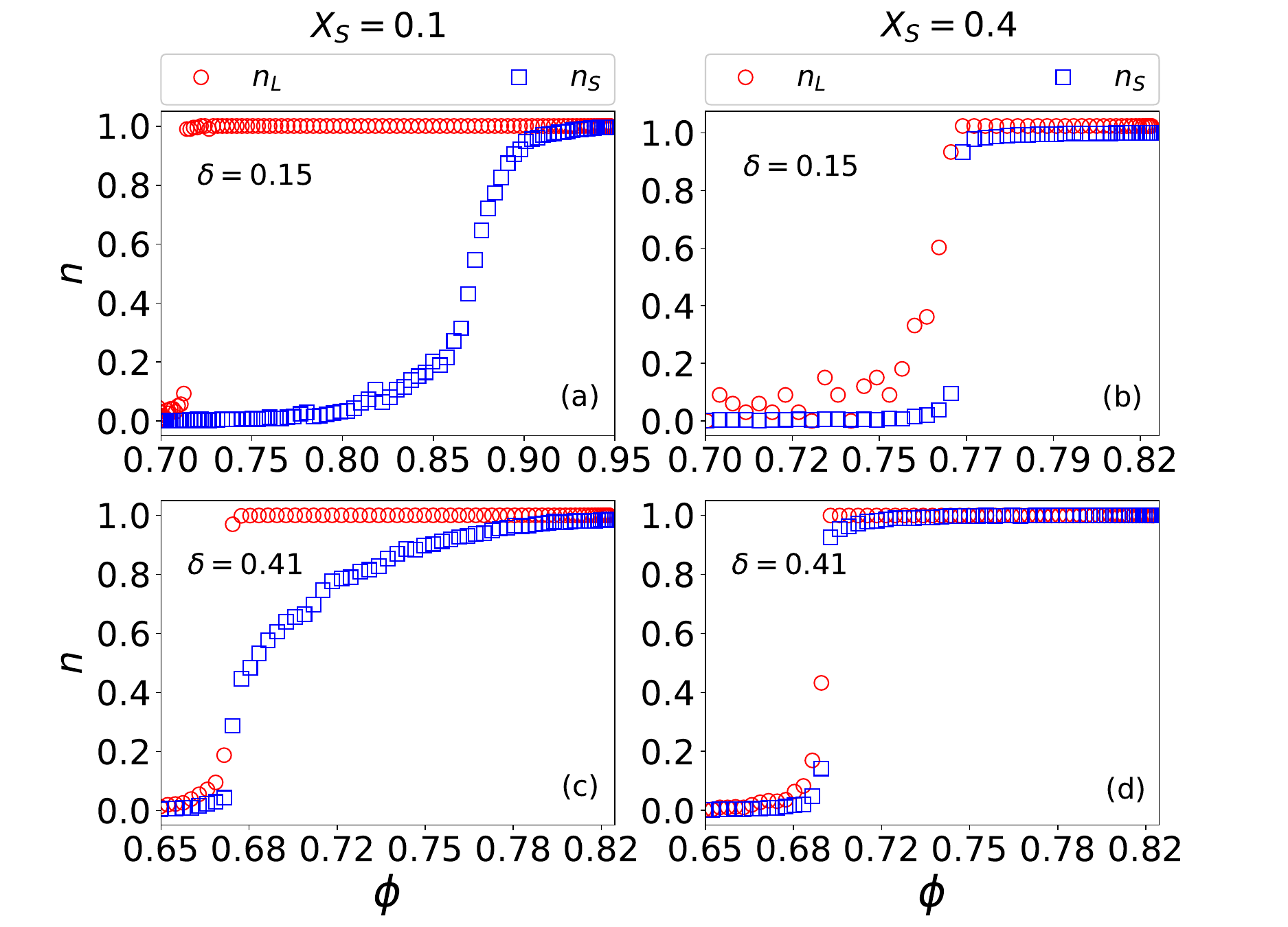}

    \caption{Fraction of large, $n_{\mathrm{L}}$, and small, $n_{\mathrm{S}}$, particles in contact, 
    as functions of the packing fraction, $\phi$, for different combinations 
    of $\delta$ and $X_{\mathrm S}$. }
	\label{fractpart_nu}
\end{figure}

\subsection{Representative example case}

Fig.~\ref{protocol} (Bottom) shows a log-scale plot of the energy ratio for three bidisperse packings 
at $X_{\mathrm S} = 0.1$. We observe that for $\delta = 0.15$, the kinetic energy 
is decaying, since collision and background medium dissipate the kinetic energy of the particles. 
Due to an applied strain, the system has a big chance to re-arrange, irreversibly, which creates 
kinetic energy  \cite{luding2021jamming}. Due to both mechanisms, the kinetic energy remains 
considerable during relaxation and compression, it decays closer zero only 
after re-arrangements and when decompression starts slowly. 
For higher $\delta$, the kinetic energy drops towards zero only later, 
at the respective jamming densities. Such differences in the behavior of the 
kinetic energy during loading and unloading (compression and decompression) are 
due to a different relaxation densities and due to the different elasticity of 
collisions between particles, e.g., $e \sim 0.95$ at higher $\delta$, see Fig.~\ref{AppA}. In this 
case, background dissipation might take place by dissipating most of the kinetic energy. After the relaxation 
period, the system is further isotropically compressed (loading) until a target maximum packing fraction 
$\phi_{\rm {max}} > \phi_{J}$ is achieved, which depends on the values of ($\delta$, $X_{\mathrm S}$) 
chosen, see Fig.~\ref{protocol} (Top). Along this process, the remaining kinetic energy of the 
system drops to near zero, $E_{k}/E_{p} \approx 10^{-6}$, suggesting the development of the jammed packing.
Plotting $E_{k}/E_{p}$ in log-scale, one can see that many things are happening when
$E_{k}/E_{p} \to 0$, see Fig.~\ref{protocol} (Bottom). Here, multiple peaks appear due to particle 
rearrangements. However, we cannot be sure if these particle rearrangements affect 
the macroscopic behavior of the system. This has started to be studied 
\citep{luding2021does, luding2021jamming}, but it is not still conclusive. After the loading process,
the isotropic decompression (unloading) starts until the initial $\phi_0$ is reached again. 
In this case, all potential energy due to overlaps is translated into kinetic energy revealing the 
unjamming behavior of the systems. Once the simulation protocol ends, the 
jamming density, see the arrows in Fig.~\ref{protocol} (Bottom), and bulk modulus of each 
bidisperse packing are determined from the decompression branch since these values are less sensitive to the 
rates of deformation \cite{goncu2010constitutive}.
Note that the open arrows (1st, lower transition) show up short before the 
kinetic energy ratio rises sharply, while the solid arrow (2nd transition)
is not accompanied by such a feature - which is due to and compatible with
our procedure and choice to report the jamming transitions on the unloading
path: the 2nd transition has mostly large particles jammed, see Sec.~\ref{SecIII}.

\section{Jamming density: first and second transition}
\label{SecIII}

\begin{figure}[t]
    
    \centering \includegraphics[scale=0.29]{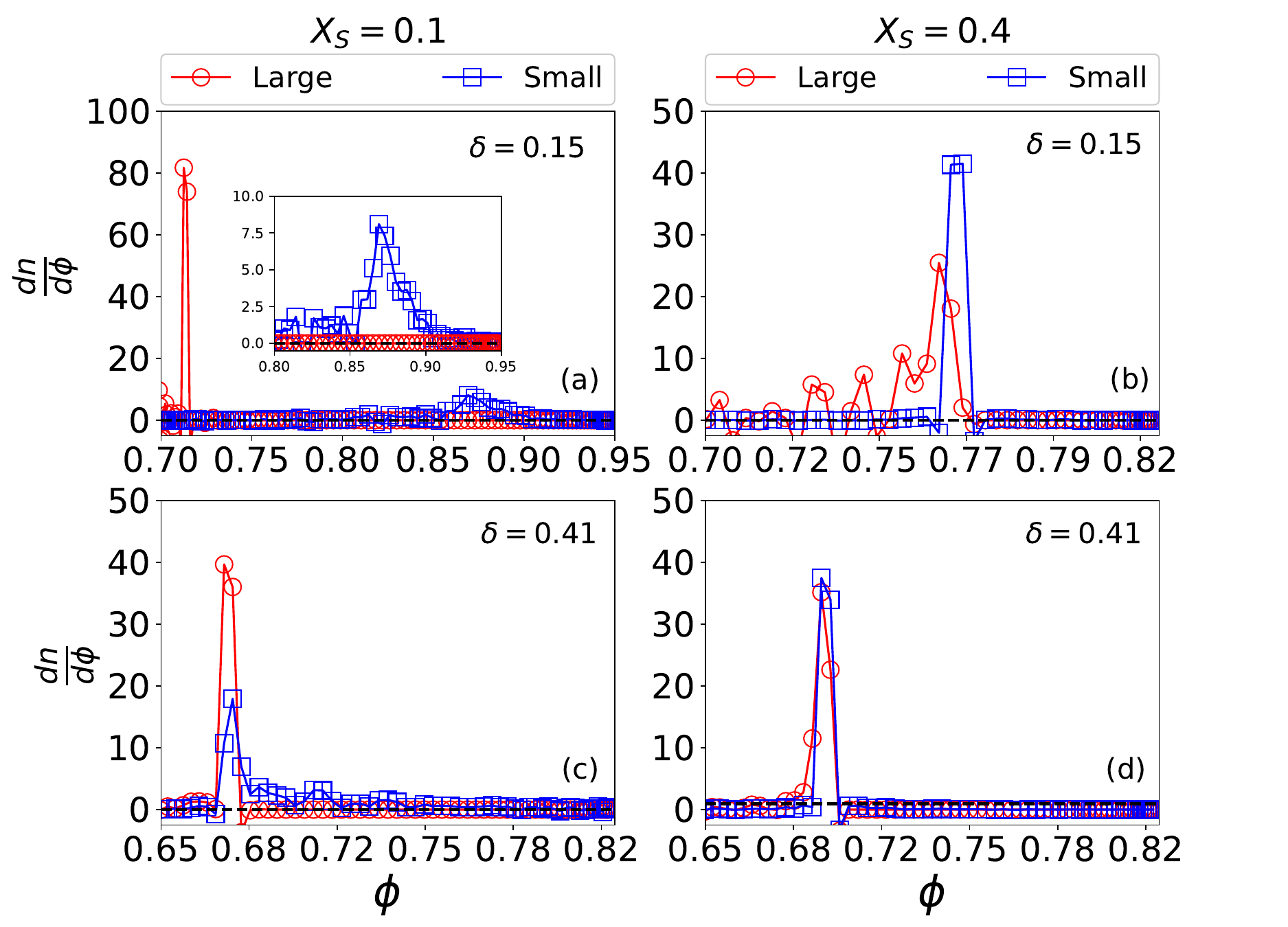}

    \caption{Derivatives of $n_{\mathrm{L}}$ and $n_{\mathrm{S}}$
    as functions of the packing fraction, for different combinations 
    of $\delta$ and $X_{\mathrm S}$. The maxi\-mum value of each derivative is 
    considered to occur at the jamming density,
    $\phi_J$. The five-point difference method with accuracy 
    $\sim O(\Delta \phi^{4})$ was used. The inset is a zoom-in 
    around the peak of the derivative of $n_{\mathrm{S}}$. 
    A clear peak is seen at $\phi \approx 0.87$, representing 
    the largest increase in the number of small particles jammed.}
	\label{Dev_fractbig_nu}
\end{figure}

We quantify the fraction of large particles, $n_{\mathrm L} = N^{c}_{\mathrm{L}}/N_{\mathrm{L}}$, 
and small particles, $n_{\mathrm S} = N^{c}_{\mathrm{S}}/N_{\mathrm{S}}$, 
contributing to the force network as a function of $\phi$ for
different $\delta$ and $X_{\mathrm S}$. 
$N^{c}=N^{c}_{\mathrm{L}}+N^{c}_{\mathrm{S}}$ is the number of 
large and small particles in contact, 
while $N_{\mathrm{L}}$ and $N_{\mathrm{S}}$ are the total number of large and small 
particles in the system. Fig.~\ref{fractpart_nu} shows jumps in $n_{\mathrm{L,S}}$ at different
$(\delta, X_{\mathrm S})$, which are equivalent to the jumps in the total mean contact 
number, see Ref.\cite{petit2020additional}.
A clear decoupling of the behavior of $n_{\mathrm{L}}$ and $n_{\mathrm{S}}$ 
is found at lower $\delta$ and lower $X_{\mathrm S}$, see Fig.~\ref{fractpart_nu} (a), while for higher 
values of $\delta$, both types of particles contribute 
simultaneously to the jammed structure, see Fig.~\ref{fractpart_nu} (b)-(d). 
Such decoupling indicates that the fraction of small particles that are jammed is
large only at higher densities, whereas large particles are jammed already at lower densities. 
To extract more precisely the value of the jamming density 
where $n_{\mathrm L}$ and $n_{\mathrm S}$ jump, 
we compute the first derivative $\partial n/ \partial \phi$ for each particle size.  
Fig.~\ref{Dev_fractbig_nu} shows the derivatives of $n_{\mathrm L}$ 
and $n_{\mathrm S}$ as functions of $\phi$, showing a characteristic peak (maximum derivative) 
at values consistent with the respective $\phi_J$. 
For $\delta = 0.15$ and $X_{\mathrm S} = 0.1$, the peak for 
large particles is found at small $\phi$, while the peak for small particles
at higher density, see Fig.~\ref{Dev_fractbig_nu} (a). 
The smaller peak amplitude is due to the smooth behavior 
of $n_{\mathrm S}$ 
\footnote{The smeared out, smoother transitions are either due to finite size,
or transients during the constant (yet small) strain-rate isotropic compression protocol,
where we minimize inertial, dynamic effects by moving slow enough to observe no 
different results for even slower rates.}
compared to $n_{\mathrm L}$, 
but still, a critical density can be extracted representing 
the largest increase in the number of small particles jammed, 
see the inset in Fig.~\ref{Dev_fractbig_nu} (a). 
This evidences that the system experiences a transition from a predominantly 
large particle structure to one where both particle sizes contribute upon compression. 
On the other hand, at higher $\delta$,  
the peaks for both fractions of particles show the same $\phi_J$ 
indicating that both particle sizes are forming the jammed structure
together, see Fig.~\ref{Dev_fractbig_nu} (b)-(d). Therefore, using this method one can extract 
the values of $\phi_J$ for the entire range of $\delta$ and $X_{\mathrm S}$.

\begin{figure}[t]
    \centering \includegraphics[scale=0.34]{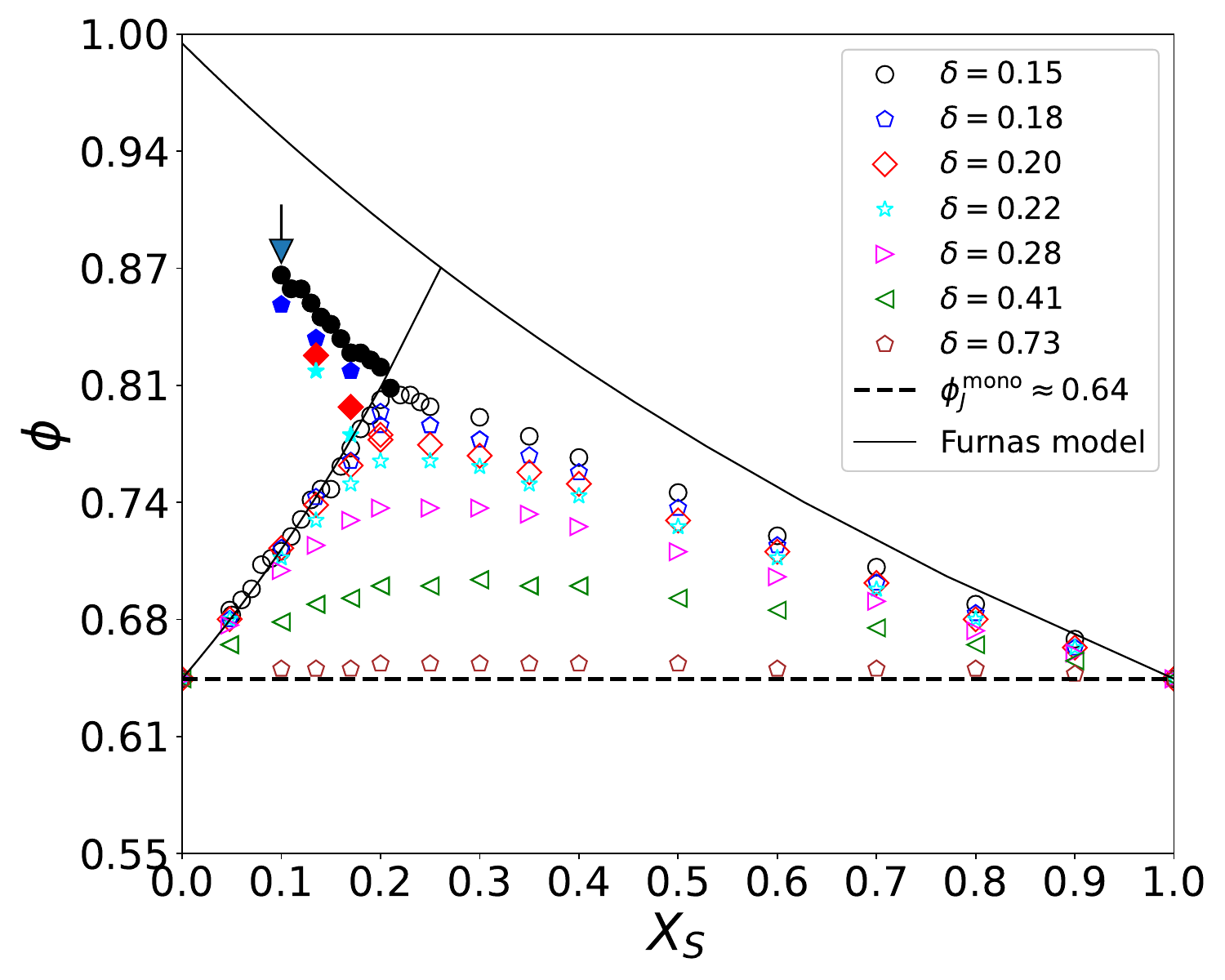}
 \caption{Jamming density, $\phi_{J}$, as a function of the concentration of small particles, 
        $X_{\mathrm S}$, for different size ratios, $\delta$. The extreme $X_{\mathrm S}$ values 
        (0 and 1) correspond to monodisperse systems, which exhibit 
        $\phi_J^{\rm mono} = \phi_{\mathrm{RCP}} \approx 0.64$, as indicated by the dashed horizontal line. 
        Solid lines represent the Furnas model of Eq.~(1) in Ref.~\cite{petit2020additional}. 
        Open (solid) symbols represent the first (second) transition line. 
        The arrow corresponds to the end-point of the second transition, $X_{\mathrm S}^{\circ}$,
        given in Fig.~\ref{fractpart_nu}, at $\phi_J \approx 0.87$.}
  \label{jamming}
\end{figure}

Fig.~\ref{jamming} displays such extracted $\phi_{J}$ values, where smaller $\delta$ 
results in jamming at higher densities for intermediate $X_{\mathrm S}$. 
Additionally, an increasing line of jamming
densities is observed for smaller $X_{\mathrm S}$. Such a line extends the transition where both 
size particles are jammed for a low range of $X_{\mathrm S}$, thus introducing a more complete jamming
diagram for bidisperse packings. This new feature of the jamming density already was 
reported in Ref.~\cite{petit2020additional}, 
where the authors made the distinction between two jammed states depending on whether small particles 
are jammed together with the large ones, or not. 
This work also showed that the second transition starts at a size ratio below around $\delta = 0.22$ 
and grows longer, extends towards smaller and smaller $X_{\rm S}$, for decreasing $\delta$. 
Comparison to an asymptotic model introduced by Furnas
in Ref.~\cite{furnas1931grading}, see Fig.~\ref{jamming},
suggests that if the size ratio of the particle types
is extreme ($\delta \to 0$), $\phi_J$ can decouple into two limit cases, 
sharing a common point at a specific $X^{*}_{\mathrm S}$.
One limit considers an approximation where large particles dominate the jammed structure, 
while small particles are not taken into account since they are too few, too small 
to play any role ($0 \leq X_{\mathrm S} < X^{*}_{\mathrm S}$). In the second limit, 
both large and small particles participate in the jammed structure ($0 \leq X_{\mathrm S} \leq 1$). 
In this case, the number of small particles is large enough to contribute and even
drive a few large particles into the jammed state.
The former limit indicates that small particles would induce contact only once 
they fill the remaining space. Therefore, the Furnas model predicts a maximum 
density of $\phi_J(X^{*}_{\mathrm S}) \approx 0.87$ at 
$X^{*}_{\mathrm S}=(1-\phi_\text{RCP})/(2-\phi_\text{RCP})\approx0.26$, 
where both line limits meet. 
This is in reasonable vicinity of the value obtained here from our 
simulations for $X^{*}_{\mathrm S} \approx 0.21$ and $\delta = 0.15$. 
In Fig.~\ref{jamming}, the second transition line given by the simulation data 
follows qualitatively the Furnas model ending at $X_{\mathrm S}^{\circ} = 0.1$ for $\delta = 0.15$. 
The transition stops at this value since for $X_{\mathrm S} < X_{\mathrm S}^{\circ}$ 
no jumps in the fraction of small particles contributing to the jammed structure are found 
(see supplementary information of Ref.~\cite{petit2020additional}). 
The additional line terminates at an end-point at some finite 
$X_{\mathrm S}^{\circ}$, which depends on $\delta$, unlike in the Furnas model.

\begin{figure}[t]
	
	\centering \includegraphics[scale=0.3]{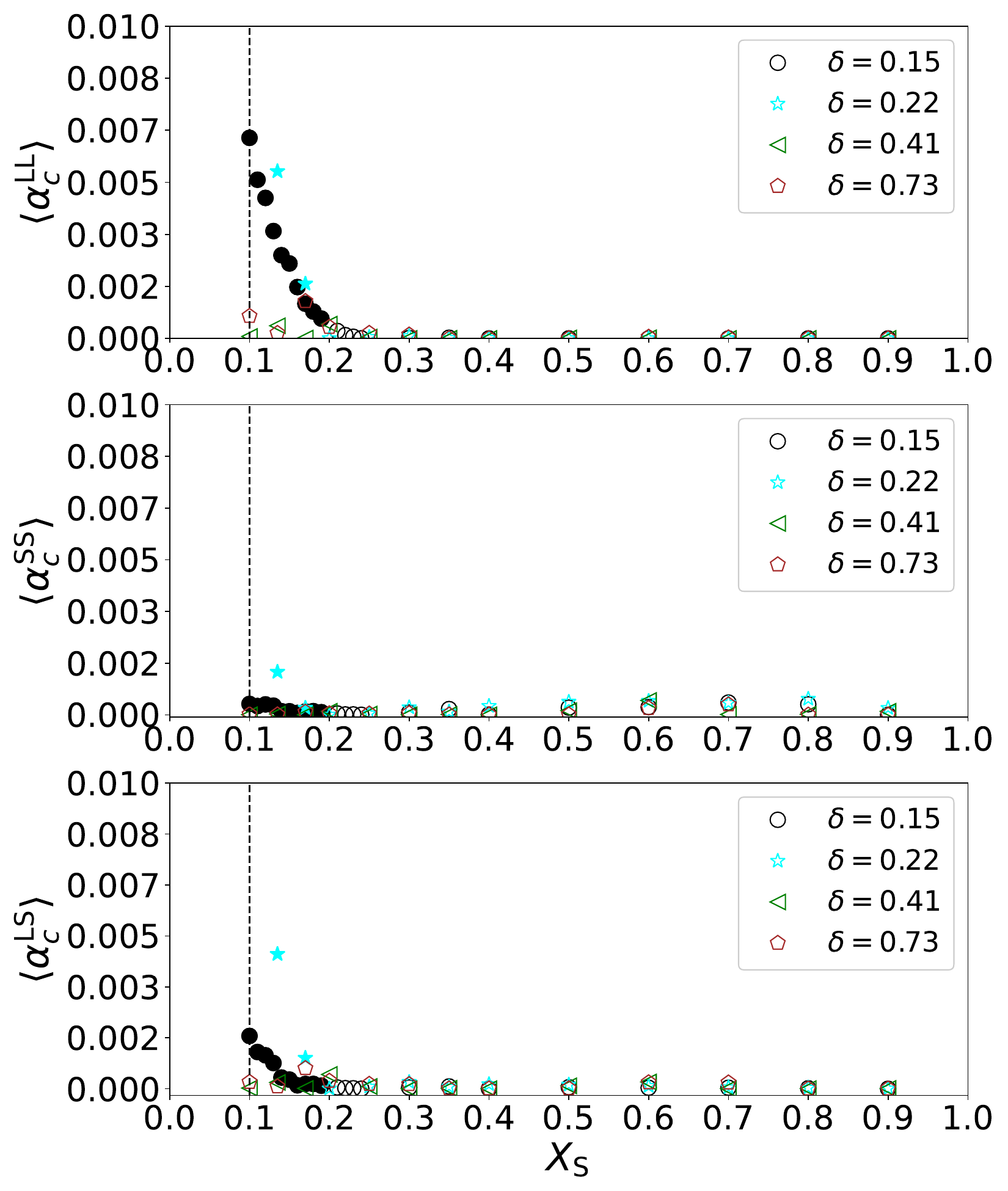}
	
	\caption{
	Dimensionless mean overlap, $\langle \alpha_c \rangle 
	= \langle \alpha'_c \rangle/ 2r'_{\mathrm{L}}$, for each contact type as a function of 
	$X_{\mathrm S}$ for different $\delta$. The values correspond to the jamming line where 
	both species of particles contribute to the jammed structure. Open (solid) symbols 
	represent the first (second) transition line. The dashed vertical line highlights 
	the endpoint found at $X_{\mathrm S} = 0.1$ for $\delta = 0.15$, represented by the 
	arrow in Fig.~\ref{jamming}. (Some noise in the data comes from the fact that we 
	are slightly above jamming, where a tiny variation may have a big effect.)}
	\label{meanover}
\end{figure}

In Ref.~\cite{suo2022unexplored} the authors have shown that
the jamming density of bidisperse packings can be lower than the monodisperse case
when $\delta \to 1$ and $X_{\mathrm{S}} \to 0$. 
This behavior is not visible in our Fig.~\ref{jamming} for $\delta = 0.73$ at
$X_{\mathrm{S}} < 0.1$ since too few data points are available in this regime that is 
not the focus of this study. 
Nevertheless, the jamming density without rattlers exhibits lower values than for the 
monodisperse case, 
see Fig.~\ref{jammingNonRattlers}.
For $\delta \to 1$, where small particles are similar to large ones, and $X_{\mathrm{S}} < 0.1$,
we expect that the few small particles more likely become rattlers inside the jammed structure 
formed by large particles, so that $\phi_J$ could decay below $\phi_{J}^{\mathrm{mono}}$. 
However, this needs to be further investigated in detail.

The high packing fractions and high pressures for which the second transitions are found,
in Fig.~\ref{jamming}, might suggest a high level of overlap (deformation) between the 
already jammed large particles in the system. 
Such extreme overlaps could lead to non-physical granular packings, since stiff real particles 
typically would break, possibly causing a much different system behavior. To quantify this, 
we determine the mean overlap, $\langle \alpha_c \rangle 
= \langle \alpha^{\prime}_c \rangle/ x'_u = \langle \alpha^{\prime}_c \rangle/ 2r'_{\mathrm{L}}$, 
for LL, SS and LS contact types along the lines where both particle species contribute to the 
jammed structure. Fig.~\ref{meanover} and appendix \ref{AppD} show 
the values of $\langle \alpha_c \rangle$ as a function of $X_{\mathrm{S}}$ at different $\delta$. 
For $\delta \geq 0.41$, the jammed packings show a mean overlap close to zero for all 
$X_{\mathrm{S}}$. For $\delta < 0.41$ and $X_{\mathrm{S}} < X^{*}_{\mathrm S}$, 
the mean overlap increases for each contact type along the second transition as a result of 
over-compression. However, the partial $\langle \alpha_c \rangle$ for the large particles is still quite low
at the second transition. The second transition presented here at low $\delta$ and low $X_{\mathrm S}$ can 
be obtained by considering experimental possibilities, an overlap around a few percent can be easily 
obtained in 3D experiments of soft PDMS sphere packings \cite{frankrichter2014disordered} 
since the maximum overlap of such a typical 
sphere can reach about $10\%$. In 2D, soft photoelastic/birefringent disks can be used as well, 
since a mean overlap of $3\%$ for $\delta = 0.71$ has been readily obtained in, e.g., 
Ref.~\cite{majmudar2007jamming}.

\section{Analysis without rattlers}
\label{SecIV}

\begin{figure}[t]
\centering \includegraphics[scale=0.31]{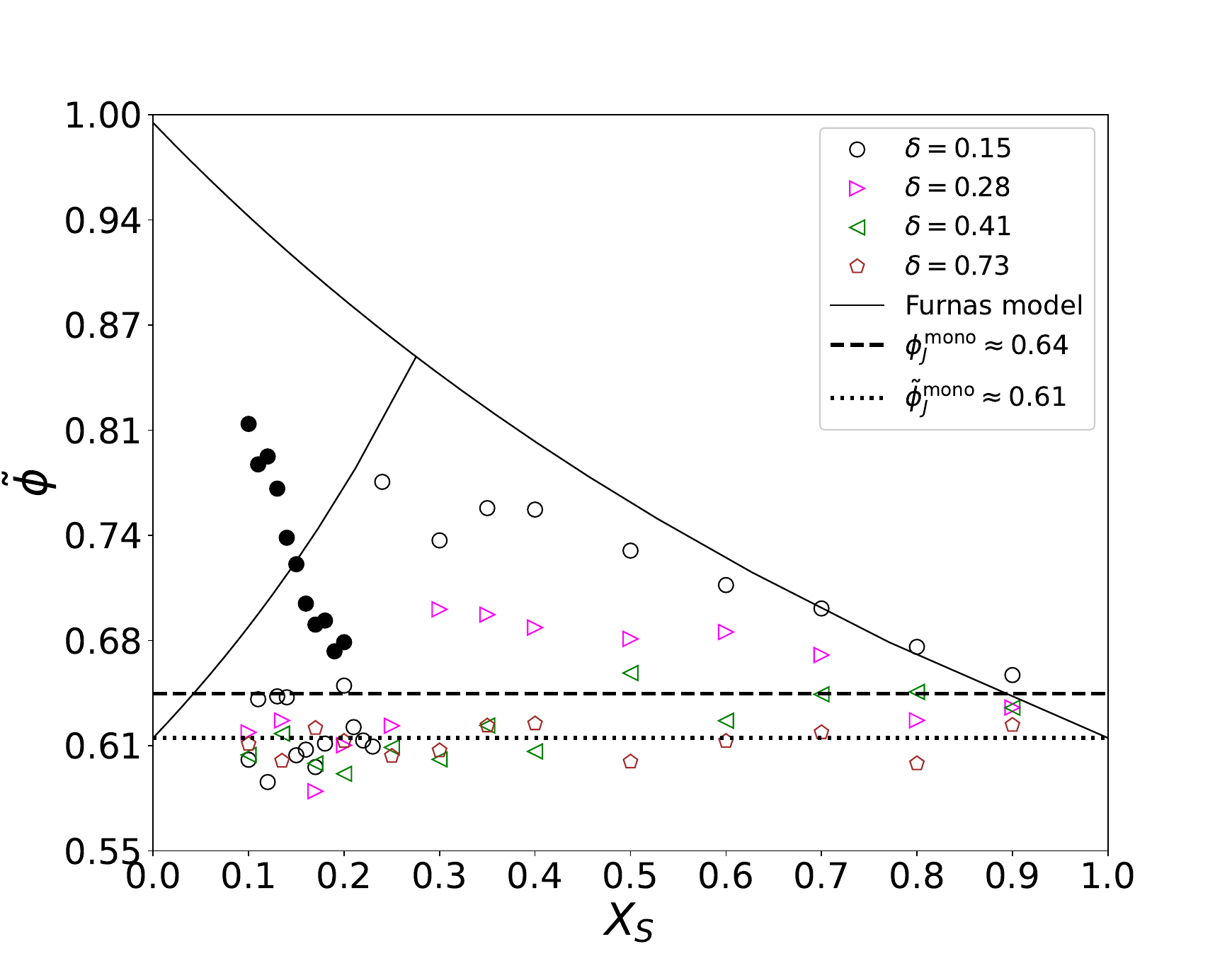}
 \caption{
 Jamming density without rattlers, $\tilde{\phi}_{J}$, as a function of the concentration of small particles, $X_{\mathrm S}$, for 
		different values of the size ratio, $\delta$. The dashed and dotted horizontal lines correspond to the jamming density 
		of monodisperse systems with ($\phi^{\rm{mono}}_J \approx 0.64$) and without rattlers
		($\tilde{\phi}^{\rm{mono}}_J \approx 0.61$). The Furnas model is 
		represented by the solid lines, see Eq.~(1) in Ref.~\cite{petit2020additional}, where $\tilde{\phi}^{\rm{mono}}_J$ was
		used as input. This shifts the model slightly downwards in the packing fraction. Open (solid) symbols represent 
		the first (second) transition line.}
\label{jammingNonRattlers}
\end{figure}

In 3D simulations of particles without friction and without gravity, rattlers are those particles 
having less than four contacts. Rattlers are not mechanically stable and do not contribute
to the force distribution \cite{kumar2016tuning, o2003jamming, agnolin2007internal_1, shaebani2008unjamming}.  
This indicates that the removal of rattlers would strongly affect variables such as the jamming densities 
found in Fig.~\ref{jamming}; how much are the mean contact number and packing 
density affected? Here, we first recalculate $\phi_J$, considering 
only those particles $i$ having four contacts or more ($\tilde{Z}_{i} \geq 4$), 
where the tilde indicates the fact that rattlers are
recursively excluded from the contact network. This means that we monitor if new particles 
with $\tilde{Z}_{i} < 4$ are created after the removal of rattlers. If so, we have to remove 
new rattlers from the contact network. This is repeated until no rattlers remain in the packing.

Fig.~\ref{jammingNonRattlers} shows the jamming density without 
rattlers, $\tilde{\phi}_{J}$, as a function of $X_{\mathrm S}$, complementing Fig.~\ref{jamming}. 
We present the same four typical size ratios, but 
similar explanations can be applied for other values of $\delta$. 
Obviously, all $\tilde{\phi}_J < \phi_J$, affecting also the monodisperse lower limit, 
$\tilde{\phi}_J^{\rm mono} \approx 0.61$. 
The jamming 
densities along the first transition line, 
between $0.1 \leq X_{\mathrm S} < X_{\mathrm S}^{*}$,
are qualitatively similar to the monodisperse case, irrespective of $\delta$, only 
lower and with some more scatter. This confirms that only large particle non-rattlers are forming 
a monodisperse jammed structure.

For all $\delta$ and $X_{\mathrm S} \geq X_{\mathrm S}^{*}$, the first transition jamming densities 
follow the Furnas' model trend, showing a similar maximum at $X_{\mathrm S}$, like in Fig.~\ref{jamming}.
However, for the smallest $\delta$, when $X_{\mathrm S} \to X_{\mathrm S}^{*} \approx 0.21$ from the right, 
$\tilde{\phi}_J$ shows a different feature of the transition. For larger $X_{\mathrm{S}}$, the system structure is formed 
by large and small particles, while for smaller $X_{\mathrm{S}}$, only large particles carry forces. 
The additional second transition of $\tilde{\phi}_J$ is also shown above the first transition, between  
$0.1 \leq X_{\mathrm S} < X_{\mathrm S}^{*}$,
as shown by the solid symbols in Fig.~\ref{jammingNonRattlers}. 
However, the jamming density of small particles on top of large particles is displaced downwards due to 
the removal of rattlers. Rather than continuing the trend from right to left, $X_{\mathrm{S}}$ decreasing, 
the second transition non-rattler jamming density drops and emerges from the monodisperse lower limit.

\begin{figure}[t]

\centering \includegraphics[scale=0.34]{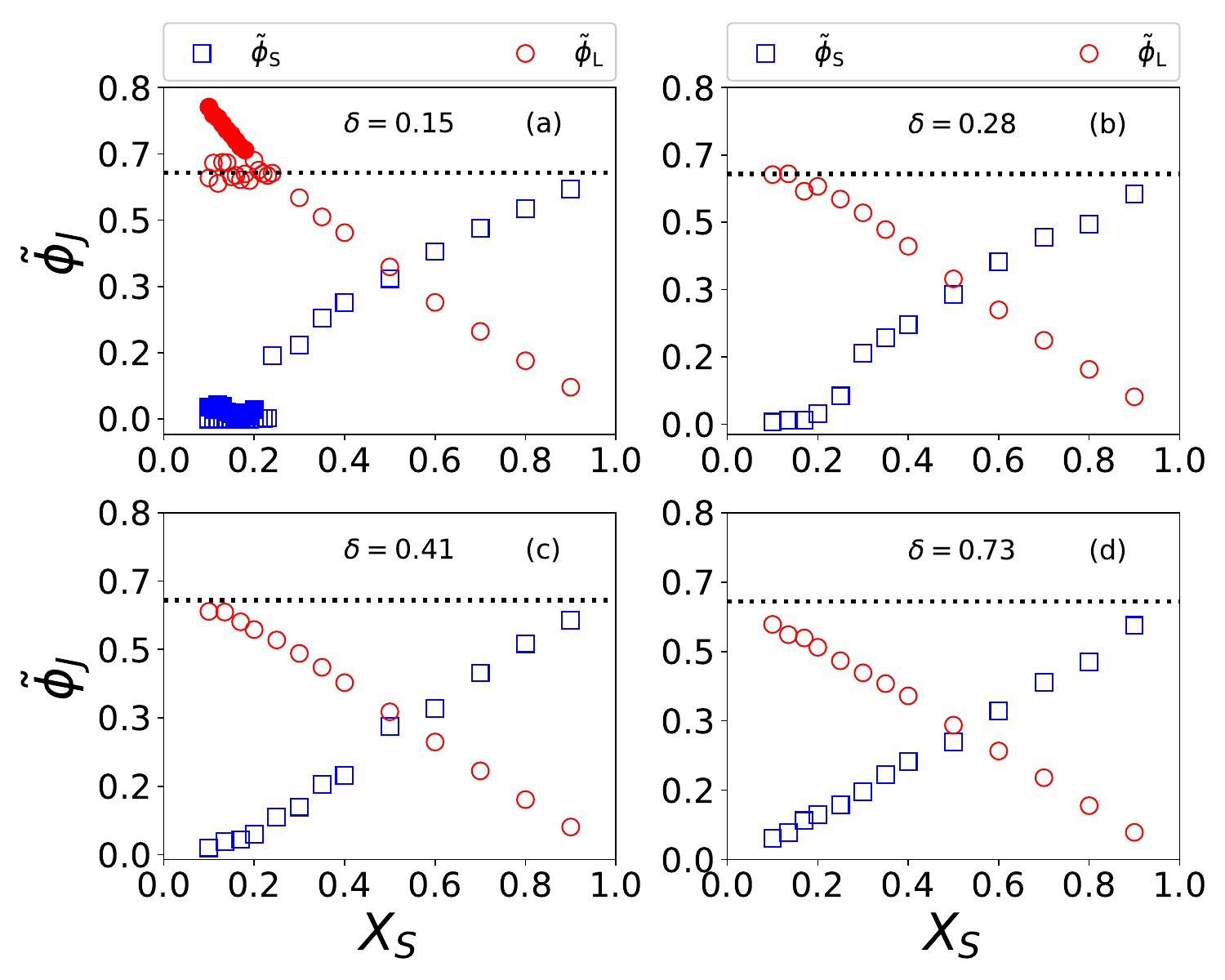}

 \caption{Partial jamming density without rattlers, $\tilde{\phi}_{\mathrm{L, S}}$, as a function of $X_{\mathrm S}$, for different 
 $\delta$. Open (solid) symbols represent the first (second) transition line. The sum 
 $\tilde{\phi}_{\mathrm{L}} + \tilde{\phi}_{\mathrm{S}} = \tilde{\phi}_{J}$ leads to the jamming densities shown in 
Fig.~\ref{jammingNonRattlers}.}
\label{Partialjamming}
\end{figure}

The jamming densities shown in Fig.~\ref{jammingNonRattlers} can be better understood by analyzing 
the partial jamming densities. 
The packing fraction 
of either large or small particles having more than four contacts, along the first and second 
transition lines. This scenario is shown in Fig.~\ref{Partialjamming}. 
For all $\delta$ along 
the first transition and between $0.1 \leq X_{\mathrm S} < X_{\mathrm S}^{*}$, 
one has $\tilde{\phi}_J = \tilde{\phi}_J^{\rm mono} \approx 0.61$ (dotted line). 
The packings behave like a monodisperse system of large particles, 
as most evident for $\delta = 0.15$, where $\tilde{\phi}_{\rm L} \approx 0.61$, 
whereas $\tilde{\phi}_{\rm S} = 0$, see Fig.~\ref{Partialjamming} (a). 
Then, $\tilde{\phi}_{\rm S}(\delta = 0.15)$ jumps to a finite value at $X_{\mathrm S} = X_{\mathrm S}^{*} \approx 0.21$, 
indicating the transition to both species contributing. For $X_{\mathrm S} > X_{\mathrm S}^{*}$, $\tilde{\phi}_{\rm S}$ 
and $\tilde{\phi}_{\rm L}$ increase or decrease, respectively.
Along the second transition, $X_{\mathrm S} < X_{\mathrm S}^{*}$, the large particles always dominate the density over small 
ones. For all $\delta$, an intersection point of equal-volume contribution is found at $X_{\mathrm S} \approx 0.5$. For instance, for 
$\delta = 0.15$, $\tilde{\phi}_{\rm L} = \tilde{\phi}_{\rm S} = 0.37$, while for $\delta = 0.73$, 
$\tilde{\phi}_{\rm L} = \tilde{\phi}_{\rm S} = 0.31$. 
Above/below the intersection point,
the small/large species dominates the first jamming of the system.

\begin{figure}[t]
\centering \includegraphics[scale=0.34]{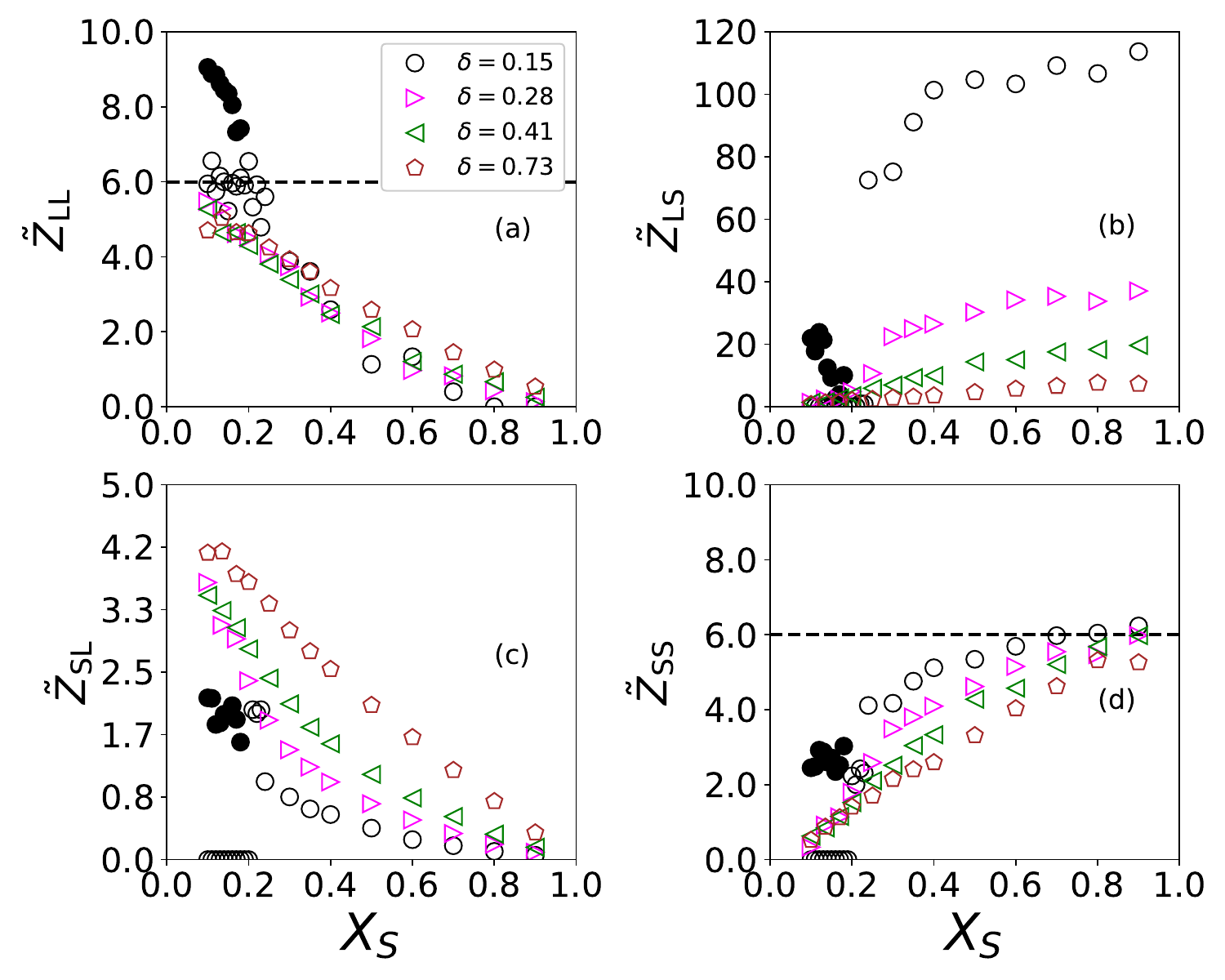}

 \caption{Partial mean contact numbers without rattlers determined at $\tilde{\phi}_J$ as a function 
 of $X_{\mathrm S}$, for different $\delta$. Open (solid) symbols represent the first (second) 
 transition line. The dashed line represents $Z_{\rm{iso}} = 6$ expected for 3D frictionless 
 monodisperse granular packings at jamming. }
  \label{partialZ}
\end{figure}

Looking deeper into the packing, we determine the partial mean contact numbers using: 
\begin{align}
\tilde{Z}_{\mathrm{LL}} &= \frac{\sum_{i = 1}^{N^{c}_{\mathrm{LL}}} \tilde{Z}^{i}_{\mathrm{LL}}}{\tilde{N}_{\mathrm{L}}} 
\label{ecu4} \\
\tilde{Z}_{\mathrm{LS}} &= \frac{\sum_{i = 1}^{N^{c}_{\mathrm{LS}}} \tilde{Z}^{i}_{\mathrm{LS}}}{\tilde{N}_{\mathrm{L}}} 
\label{ecu3} \\
\tilde{Z}_{\mathrm{SS}} &= \frac{\sum_{i = 1}^{N^{c}_{\mathrm{SS}}} \tilde{Z}^{i}_{\mathrm{SS}}}{\tilde{N}_{\mathrm{S}}}
\label{ecu5} \\
\tilde{Z}_{\mathrm{SL}} &= \frac{\sum_{i = 1}^{N^{c}_{\mathrm{SL}}} \tilde{Z}^{i}_{\mathrm{SL}}}{\tilde{N}_{\mathrm{S}}} 
\label{ecu6} 
\end{align}

where each sum is running over the number of contacts (LL, LS, SS, SL) in the packing respectively.
We have divided by $\tilde{N}_{\mathrm{L}}$ and $\tilde{N}_{\mathrm{S}}$ to examine the contribution 
of each contact type to the total coordination, $Z=Z_\mathrm{L}+Z_\mathrm{S}$ (with $Z \geq 4$). 
For instance, at large $X_{\mathrm{S}}$ the few large particles in the system have particularly large 
$\tilde{Z}_{\mathrm{LS}}$, since they are surrounded by many small ones. At the same time, 
$\tilde{Z}_{\mathrm{SL}}$ is near zero since each small particle is in contact with very few or no 
large ones. As expected, not shown, ${Z}_{\mathrm{LS}} = {Z}_{\mathrm{SL}}$ if in Eqs.~(\ref{ecu3}) 
and (\ref{ecu5}) the denominator is replaced by the total 
$\tilde{N} = \tilde{N}_{\mathrm{L}} + \tilde{N}_{\mathrm{S}}$.

Fig.~\ref{partialZ} (a) shows $\tilde{Z}_{\rm{LL}} \approx Z_{\mathrm{iso}} = 6$
for $\delta = 0.15$ between $0.1 \leq X_{\mathrm S} < X_{\mathrm S}^{*}$,
while $\tilde{Z}_{\rm{SS}}$ remains zero (open circles). This is one of the reasons 
that both jamming density and bulk modulus have 
approximately constant values in this regime of $X_{\mathrm S}$, 
see Fig.~\ref{jamming} and Fig.~\ref{BMjmming}, respectively. 
For higher $X_{\mathrm S} > X_{\mathrm S}^{*}$, 
$\tilde{Z}_{\mathrm{LL}}$ and 
$\tilde{Z}_{\mathrm{SL}}$ decay to zero while $\tilde{Z}_{\mathrm{SS}}$ and $\tilde{Z}_{\mathrm{LS}}$ increase. 
Note that $\tilde{Z}_{\mathrm{SS}} \approx Z_{\mathrm{iso}}$ as $X_{\mathrm S} \to 1$, since 
the jammed packing is formed by small particles only.

Along the second transition line, 
$\tilde{Z}_{\mathrm{LL}}$ and
$\tilde{Z}_{\mathrm{LS}}$ 
decrease within $ 0.1 \leq X_{\mathrm S} < X_{\mathrm S}^{*}$. 
Note that $\tilde{Z}_{\mathrm{LL}} > \tilde{Z}_{\mathrm{iso}}$ along the second transition since the 
packing is becoming denser so that the overlaps and the contact numbers between large particles 
increase with decreasing $X_{\mathrm{S}}$. 
The mixed contacts of large particles, $\tilde{Z}_{\mathrm{LS}}$, show many contacts with small 
ones for $\delta=0.15$
with a dip at $X_{\mathrm S} \approx X_{\mathrm S}^{*}$. In contrast, the contacts related to 
small particles display a maximum at $X_{\mathrm S} \le X_{\mathrm S}^{*}$.

The results presented in Fig.~\ref{partialZ} show how differently the two 
particle species in the bidisperse granular packing contribute to its rigid structure along 
the first and second jamming transition. Similar values of the partial mean contact number along the 
first jamming transition were reported in Ref.~\cite{biazzo2009theory}, disregarding the second transition.

\section{Bulk modulus near the jamming density}
\label{SecV}

\begin{figure}[t]
  \centering \includegraphics[scale=0.3]{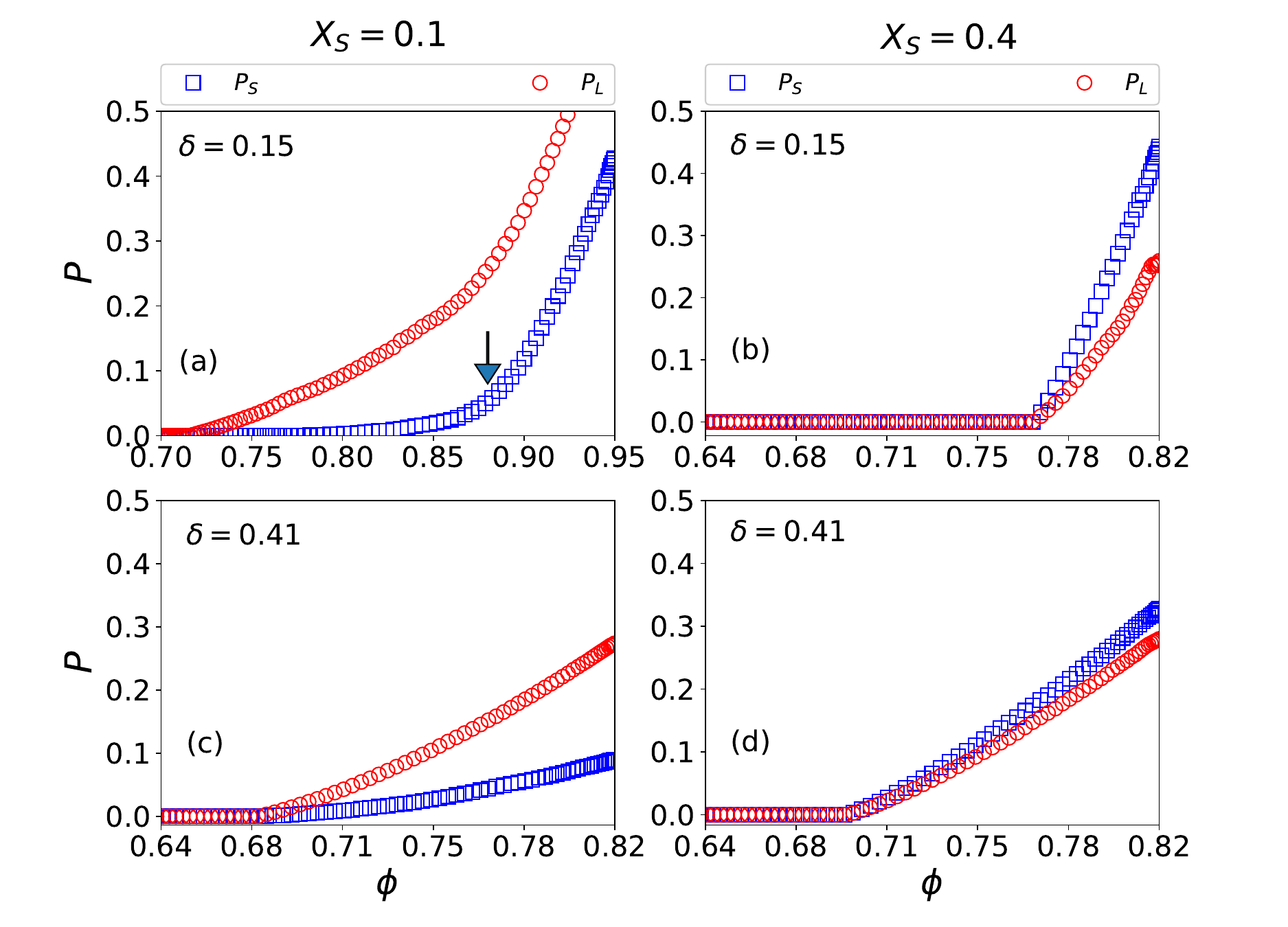}
 \caption{ 
 Partial dimensionless pressures, $P$, 
 as a function of $\phi$ for four combinations of $\delta$ and $X_{\mathrm S}$. 
 Note the different horizontal axis scaling top-left and the extreme overlaps (deformations) 
 at large $\phi$. The arrow indicates the second transition, $\phi_J \approx 0.87$, 
 extracted by 
 the derivative of $n_{\mathrm{S}}$, see Fig.~\ref{fractpart_nu} and Fig.~\ref{Dev_fractbig_nu}. }
  \label{pressure}
\end{figure}

Here we focus on the pressure of each particle species since they are
the most important variable to determine how much they contribute to the force network and thus to the
bulk modulus. Therefore, rattlers are considered since they can also contribute to 
the stiffness of the jammed packing upon compression. Appendix~\ref{AppB} shows the derivation of the large and 
small dimensionless pressures: 
$P_{\mathrm{L}} = 2 r'_{\mathrm{L}} P'_{\mathrm{L}}/\kappa'_{n}$ and 
$P_{\mathrm{S}} = 2 r'_{\mathrm{L}} P'_{\mathrm{S}}/\kappa'_{n}$, respectively.

Fig.~\ref{pressure} shows typical evolutions of the dimensionless pressures of each 
particle size with packing fraction, $\phi$, for four combinations of 
$\delta$ and $X_{\mathrm S}$. Irrespective of $X_{\mathrm S}$, $P_{\mathrm{L}}$ and $P_{\mathrm{S}}$ 
show values of zero at very low $\phi$ since the system is not jammed. 
When the system becomes jammed, $P_{\mathrm{L}}$ and $P_{\mathrm{S}}$ start showing non-zero 
values for $\phi > \phi_J$. For $X_{\mathrm S} = 0.4$, Figs.~\ref{pressure} (b,d) show that 
each type of $P$ starts rising simultaneously, both contributing to the jammed structure, 
at $\phi_J \approx 0.77$ (for $\delta = 0.15$) 
and $\phi_J \approx 0.70$ (for $\delta = 0.41$). 
For $X_{\mathrm S} = 0.1$, the values of $P$ 
for $\delta = 0.41$ show a similar behavior to those found for $X_{\mathrm S} = 0.4$, see 
Fig.~\ref{pressure} (c). However, reducing the size ratio until $\delta = 0.15$, 
$P$ shows a different scenario, see Fig.~\ref{pressure} (a). In this case, the jamming of the system is 
initially driven by the jammed structure formed by large particles at $\phi_J \approx 0.71$. 
At this value, the small particles are located in the cages formed by the large ones having 
zero contacts and contributing zero partial pressure, as was previously indicated in Refs.~\cite{kumar2016tuning, 
petit2020additional}. This scenario, where predominantly large particles are jammed, has been 
widely considered in the literature as the jamming state of a bidisperse system but it is 
the jamming state of a monodisperse packing of large particles with only a few contacts among small 
particles carrying load \cite{kumar2016tuning}. Making the packing denser, some small 
particles make contact with the structure of large particles revealing a continuous 
rising but low value of $P_{\mathrm{S}} \ll P_{\mathrm{L}}$ driven mainly by LS contacts particles (not shown explicitly here), 
see the weak increase of $n_{\mathrm{S}}$ and $P_{\mathrm{S}}$ around $\phi_J \approx 0.76$ 
given in Fig.~\ref{fractpart_nu} and Fig.~\ref{pressure} (a). Then, at higher densities a 
large number of small particles quantified by $n_{\mathrm{S}}$ get jammed at $\phi_J \approx 0.87$ 
contributing considerably $P_{\mathrm{S}} < P_{\mathrm{L}}$. 
In this case, both SS and SL contacts contribute 
to $P_{\mathrm{S}}$, see Fig.~\ref{partialvalues} (a). Similar
behaviors are obtained when looking at the partial mean contact number 
where a discontinuity at jamming is also found, see Ref.~\cite{petit2020additional}.

\begin{figure}[t]
    \centering \includegraphics[scale=0.3]{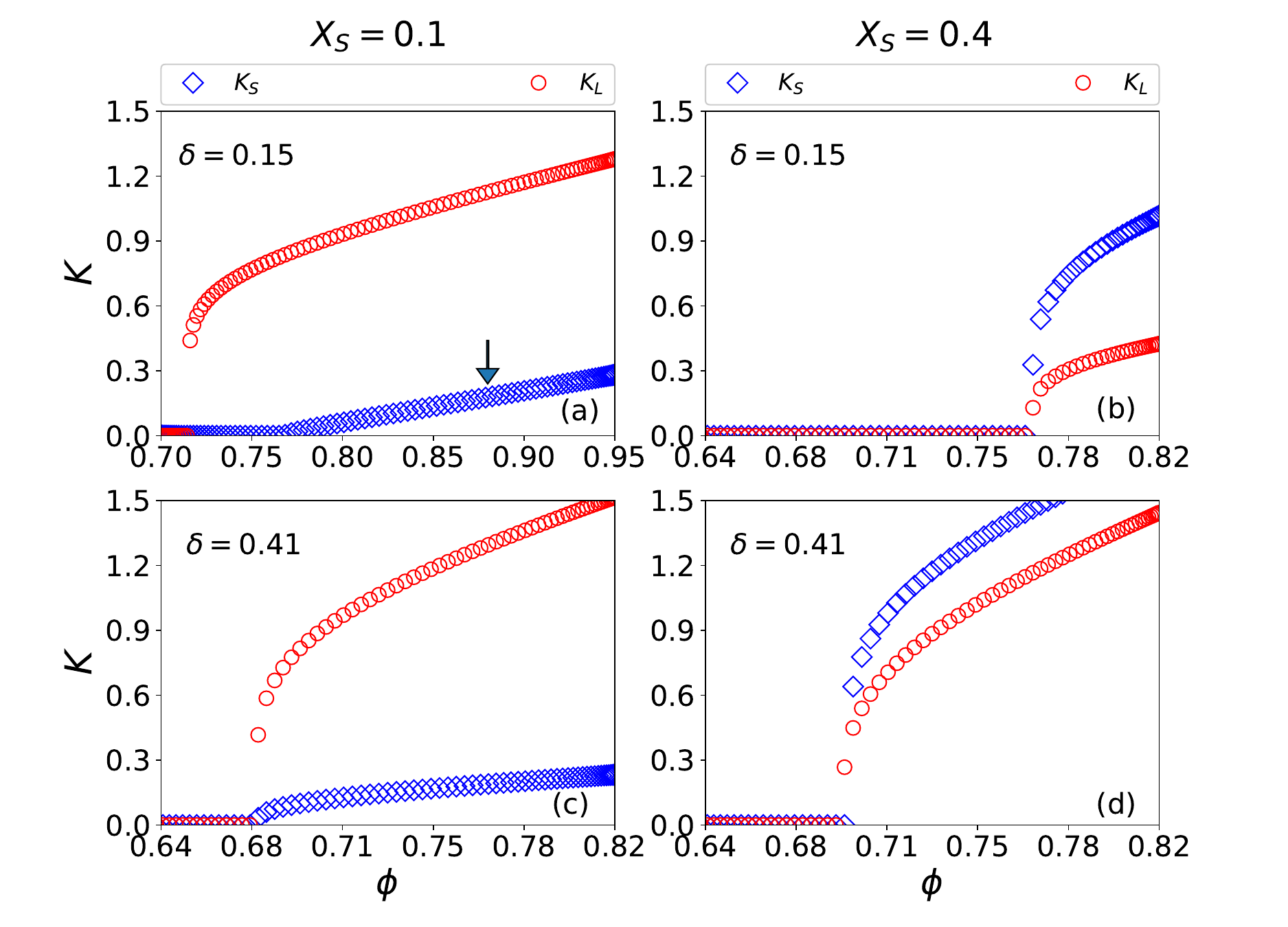}
 \caption{ Dimensionless bulk modulus, $K$, as a function of $\phi$ for four 
 combination of $\delta$ and $X_{\mathrm S}$. The arrow indicates the
 second transition, $\phi_J \approx 0.87$, shown in Fig.~\ref{fractpart_nu}.}
  \label{BulkM}
\end{figure}

\begin{figure}[t]
\centering \includegraphics[scale=0.33]{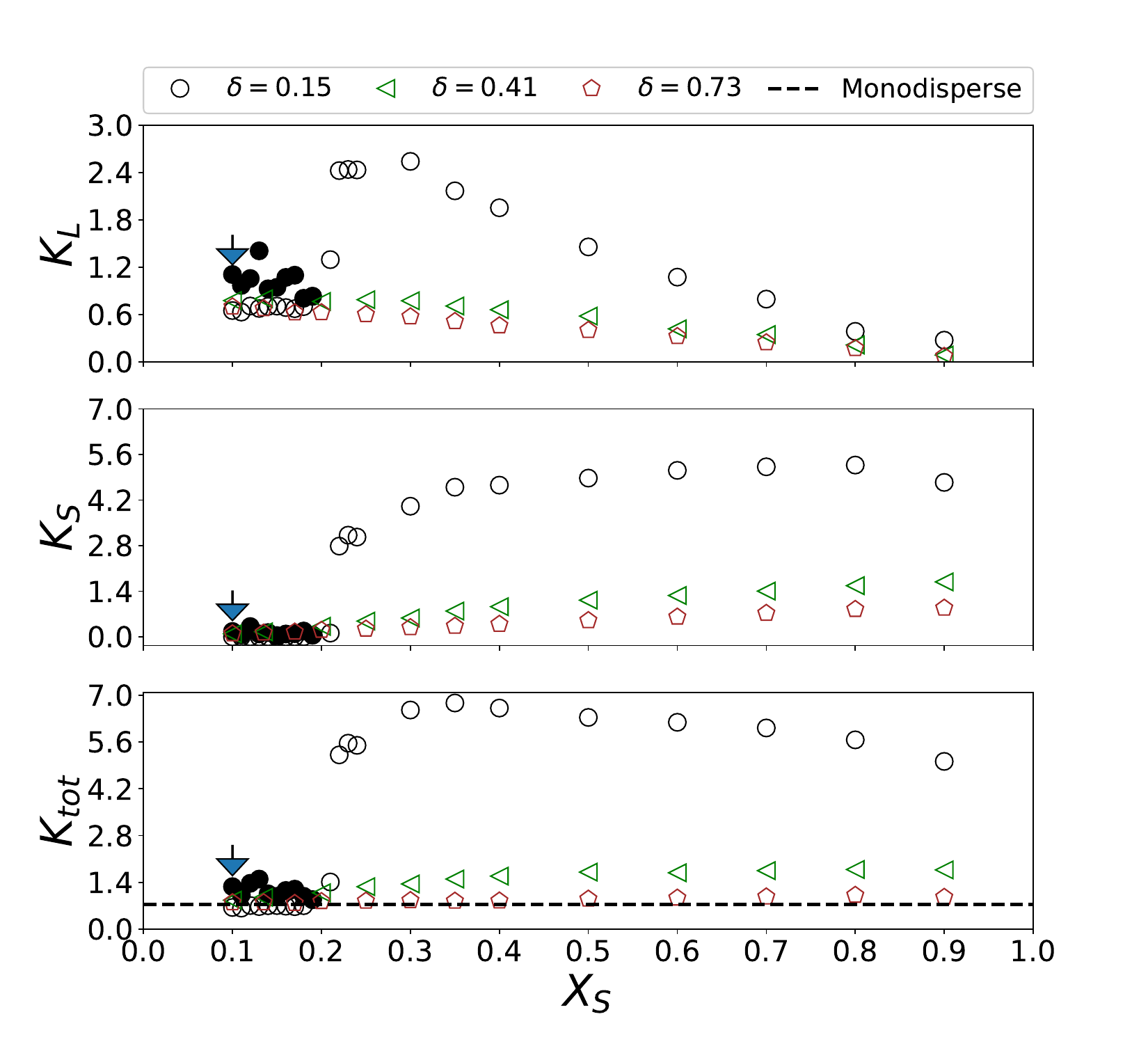}
\caption{Partial and total dimensionless bulk modulus, $K$, as a function of $X_{\mathrm S}$ for different 
$\delta$. Open (solid) symbols represent the first (second) transition. 
These $K$ values were obtained roughly $1\%$ above the jamming density $\phi_J$ shown in Fig.~\ref{jamming},
which causes the ambiguous scatter.
The dashed line corresponds to the total dimensionless bulk modulus for monodisperse packings, 
$K_{\mathrm{mono}} \approx 0.76$. Note the different vertical axis scaling in the top panel.
The arrow indicates the second transition shown in Fig.~\ref{fractpart_nu} (a), $\phi_J \approx 0.87$.}
\label{BMjmming}
\end{figure}

Each of the jamming density values displayed in Fig.~\ref{jamming} represents 
a jammed structure that can show resistance when the external stress is applied. 
The system behaves like an elastic solid. One of the properties 
that quantify the mechanical behavior of granular packings is the bulk modulus, $K= \phi\,\partial P /\partial \phi$,  
the change of pressure, $P$, with packing fraction, $\phi$. 
To determine the dimensionless bulk moduli, $K_{\mathrm{L}}$ and $K_{\mathrm{S}}$, as 
a function of $\phi$, we use the empirical fit-equation $P = P_{0}(\phi - \phi_{J})^{a}$ to approximate
the dimensionless pressures, $P_{\mathrm{L}}$ and $P_{\mathrm{S}}$.
This is consistent with 
Ref.~\cite{petit2020additional}, where one jamming density was determined for each particle size.
In previous works \citep{majmudar2007jamming, aharonov1999rigidity} 
a value of $a \approx 1.1$ for $\delta = 0.71$ was reported. We find here that the power
$a$ depends on the pressure, for each particle size, and on the combination ($\delta$, $X_{\mathrm S}$).
In particular, we find that for high $\delta$ the fitting parameters of 
the power law do not change too much with the fitting range, thus the fitting range 
is arbitrary, see Fig.~\ref{param_SRp4143}. However, at low $\delta$ the 
fitting parameters for $P_{\mathrm{S}}$ are quite sensitive to the fitting range while those related to 
$P_{\mathrm{L}}$ only slightly vary, see Fig.~\ref{param_SRp1548}. 
We think that such 
variations of $P_{\mathrm{S}}$ are caused by the high overlaps developed at high $\phi$, 
which tend to modify the fitting parameters. Therefore, we fit $P_{\mathrm{S}}$ in 
smaller ranges, near to $\phi_J$, keeping $a_{\mathrm{S}}$ as close as possible to $a_{\mathrm{L}}$. 
This results in low overlaps and similar power laws between large and small particle pressures. 
Full details of the fitting parameters are given in Appendix~\ref{AppC}.

Once the fitting parameters for $P_{\mathrm{L}}$ and $P_{\mathrm{S}}$ are extracted, 
one can determine the dimensionless bulk modu\-li of large and small particles using their definition:
$K = \phi\,P_{0}\,a\,(\phi - \phi_{J})^{a-1}$. Fig.~\ref{BulkM} shows the partial dimensionless bulk 
moduli as a function of $\phi$, for the same combinations of $(\delta, X_{\mathrm S})$ as in 
Fig.~\ref{pressure}. 
We can see a jump, similar to those observed in Fig.~\ref{fractpart_nu}, 
for both particle sizes at
 high $\delta$ and high $X_{\mathrm S}$, see Fig.~\ref{BulkM} (b)-(d), least pronounced in (c). 
However, Fig.~\ref{BulkM} (a) shows two behaviors of $K_{\mathrm{L}}$ and $K_{\mathrm{S}}$, jamming at 
different $\phi$, with and without a jump, respectively. 
This confirms that at lower densities large particles jam first, forming a structure with an overall bulk modulus.
Then, at $\phi \approx 0.76$, 
some small particles make contact with larges ones, leading to $K_{\mathrm{S}} \neq 0$.
Increasing $\phi$, small particles make gradually more contact with large ones until a considerable 
amount of small particles make many contacts also with those small particles already jammed, see the arrow
in Fig.~\ref{BulkM} (a) and the jump in Fig.~\ref{fractpart_nu} (a). Although, no jump in $K_{\mathrm{S}}$ 
is obtained here, a jump in the bulk modulus between small-small particles, $K_{\mathrm{SS}}$, is indeed observed, 
see Fig.~\ref{partialvalues} (e). Such jump is hidden by the bulk modulus given by the mixed contacts, 
$K_{\mathrm{LS}}=K_{\mathrm{SL}}$, since it dominates $K_{\mathrm{S}}$. 
The second transition where small particles are jammed jointly with large ones indicates a 
different packing structure with a higher bulk modulus according to $K = K_{\mathrm{L}} + K_{\mathrm{S}}$. For 
the case of the force network, large 
particles carry high forces while small ones carry low forces, as has been presented in 
Refs.~\cite{petit2017contact, petit2018reduction}, 
and is confirmed by our overlap data shown above in Fig.~\ref{meanover}. 
However, it would be interesting to study how the 
force distribution changes according to size ratio and concentration of small particles, 
especially for lower size ratios, where the second transition appears. The results shown in Fig.~\ref{BulkM} 
demonstrate that for low $\delta$ and $X_{\mathrm S}$, small particles jam only at large 
$\phi$ to provide a different jammed structure and as consequence a higher bulk modulus.

Next, we extract the dimensionless bulk modulus at a $\phi$ slightly above each jamming density
shown in Fig.~\ref{jamming}. As an arbitrary choice, we evaluate the 
data when $\phi-\phi_J$ reaches $1\%$ above each $\phi_J$.
Fig.~\ref{BMjmming} shows the variation 
of the partial and total $K$ as a function of $X_{\mathrm S}$ for different values of $\delta$. 
For $\delta \geq 0.41 $ and low $X_{\mathrm S}$, large particles dominate the structural resistance 
over small ones, $K_{\mathrm{L}} >  K_{\mathrm{S}}$. At larger $X_{\mathrm S}$, $K_{\mathrm{L}}$ 
is reduced while $K_{\mathrm{S}}$ increases showing an intersection point at a specific $X_{\mathrm S}(\delta)$. 
This point marks the onset of domination of the small particles in the structure over the large ones, 
$K_{\mathrm{S}} \ge K_{\mathrm{L}}$. 
The intersection point between 
$K_{\mathrm{L}}$ and $K_{\mathrm{S}}$ has nothing to do with the intersection point shown in Fig.~\ref{packings}. 
In the traditional 50:50 mixture cases, 
the bulk modulus is not equally distributed, but is dominated by the large particles.

Interestingly, the total dimensionless bulk modulus for the larger $\delta \ge 0.41 $ is largely
independent of $X_{\mathrm S}$ showing almost constant values, almost the same as monodisperse packings for: 
$K_{\mathrm{tot}}(\delta = 0.73) \approx 0.76$. 
For even smaller $\delta = 0.15$, a different behavior is found. Along the first 
transition (open symbols), $K_{\mathrm{L}}$ shows a constant value for 
$0.1 \leq X_{\mathrm S} < X_{\mathrm S}^{*}$, while $K_{\mathrm{S}} = 0$, indicating 
that predominantly large particles provide the resistance of the system. 
At $X_{\mathrm S}^{*} \approx 0.21$, both $K_{\mathrm{L}}$ and  $K_{\mathrm{S}}$ show a jump to a specific value, 
which is consistent with the peak in $\phi_{J}$ shown in Fig.~\ref{jamming}. 
For $X_{\mathrm S} > X_{\mathrm S}^{*}$, $K_{\mathrm{L}}$ decays to zero, 
whereas $K_{\mathrm{S}}$ 
increases until 
the small particles dominate the structural 
resistance of the system. 
The total dimensionless bulk modulus, $K_{\mathrm{tot}}$, shows a similar 
jump at $X_{\mathrm S}^{*}$, exhibiting its highest value at $X_{\mathrm S} \approx 0.35$, to then 
tending to the monodisperse value as $X_{\mathrm S} \to 1$.

As was demonstrated in Ref.~\cite{petit2020additional} and shown in Fig.~\ref{jamming}, 
small particles, indeed, begin to strongly contribute to the jammed structure already formed by 
large particles -- as indicated by the jump in $K$ at the second transition. This
makes the system stiffen, showing a second transition line at low $X_{\mathrm{S}}$ and low $\delta$, 
see solid symbols in Fig.~\ref{BMjmming}. Along the second transition line, $K_{\mathrm{S}}$ indeed 
increases within $X_{\mathrm{S}} \in [0.1,0.21]$ contrary to the zero values found along the 
first transition. In this range, $K_{\mathrm{S}}$ increases while $K_{\mathrm{L}}$ decreases with increasing 
$X_{\mathrm{S}}$ towards $K_{\mathrm{tot}} \approx K_{\mathrm{mono}}$. 
The gap between the first and second transition obtained at low $X_{\mathrm{S}}$ 
for $K_{\mathrm{tot}}$ demonstrates that the structural resistance increases by around $50\%$ when 
small particles contribute to the jammed structure of the system.

In a very recent paper, similar results of the bulk modulus as a function of $X_{\mathrm S}$  
were shown \cite{hara2021phase}. 
The authors found jump-like behavior for $\delta = 0.17$ near $X_{\mathrm S} \approx 0.21$ for 
very low pressure levels,
indicating the first jamming transition. While the authors did not explicitly identify
the additional transition line shown in Fig.~\ref{jamming}, their results suggest that
the small particles can cause the second transition at higher pressure levels.

\section{Summary, Conclusions and Outlook}
\label{SecVII}

In summary, we show that in bidisperse packings of soft particles, 
the dimensionless bulk modulus, $K$, can show both sharp and smooth transitions at 
first or second transition depending on which parameters are changed.
Its magnitude depends on whether large and small particles 
jam simultaneously or not.
We find a critical combination of size ratio and concentration of small particles, 
$(\delta_{\mathrm c} = 0.22, X^{*}_{\mathrm S} \approx 0.21)$,
below which the jamming density and bulk modulus present a new, additional transition,
mostly disregarded, except in some previous literature \cite{kumar2016tuning,petit2020additional}.
For $\delta > \delta_{\mathrm c}$ and $X_{\mathrm S} > X^{*}_{\mathrm S}$, 
the bulk modulus comes from a structure formed by both large and small particles that simultaneously 
jam at the same density. 
However, for $\delta < \delta_{\mathrm c}$ and $X_{\mathrm S} < X^{*}_{\mathrm S}$, 
the bulk modulus displays two transitions, obtained at different densities that depend on 
particle size. This means that large particles first jam at low densities forming an initial 
structure where small ones do mostly not contribute. It is only at higher densities that the
small particles start to get jammed jointly with large particles forming a  
different structure.

The highest bulk modulus was observed at $X^{K_{\mathrm{max}}}_{\mathrm{S}} \approx 0.35$, for the lowest size ratio 
considered, $\delta = 0.15$, which is far from the highest jamming density shown in the 
jamming diagrams in Fig.~\ref{jamming} and Fig.~\ref{jammingNonRattlers}.
Although many small particles are jammed below $(\delta_{\mathrm c}, X_{\mathrm S}<X^{*}_{\mathrm S})$, 
the total bulk modulus shows a much lower magnitude compared to the bulk modulus above  
$(\delta_{\mathrm c}, X_{\mathrm S} > X^{*}_{\mathrm S})$, see Fig.~\ref{BMjmming}. 
Indeed, for $\delta = 0.15$, $K$ 
increases by approximately $50\%$ to the case when small particles are jammed jointly 
with large ones at a higher density. This suggests that distinct jammed structures are created at first 
and second transitions, where much stiffer structures are created for larger $X_{\mathrm{S}}$.

The behavior of the bulk modulus, and its relation to the jamming density,
as presented along the first and second jamming transitions, have given better insights
into the mechanical properties of jammed bidisperse systems. 
By tuning the values of $\delta$ and $X_{\mathrm S}$, one can get stiffer or less stiff 
bidisperse structures. The different values of the bulk modulus obtained here, suggest
that other properties such as shear modulus, vibrational density of states,
or force distributions must be different as well.

Future research could aim at understanding how $\phi_J$ changes when different materials
(different restitution coefficients, $e$, dissipation by contact damping, $\gamma_{n}$, 
or background damping, $\gamma_{b}$), different contact friction coefficients, $\mu_p>0$, 
different size- and shape-distributions, or different testing protocols (compression rates) are used. 
This would give us a broader overview of the dependence of the jamming density on particle 
properties, the surrounding medium, and also protocol/method and modes of deformation (isotropic
vs.\ shear/deviatoric). This might eventually allow understanding the interplay of a fluid with the 
particles, when (a) small, fines are washed out of a matrix formed by jammed 
larger particles, or (b) the dynamics and statics of suspended particles, 
possibly display a crossover from a viscosity dominated, slow regime to an 
inertial dominated, more rapid flow state, where the large and small particles
interact with the fluid in strongly different ways.

\section*{Acknowledgments}

We thank Till Kranz and Peidong Yu
for helpful discussions. This work was supported by the German Academic Exchange Service (DAAD)
under the grant n$^{\circ}$ 57424730.


\appendix

\section{Restitution coeff. vs. size-ratio}
\label{AppA}

\begin{figure}[htbp]
  \centering
    \includegraphics[scale=0.425]{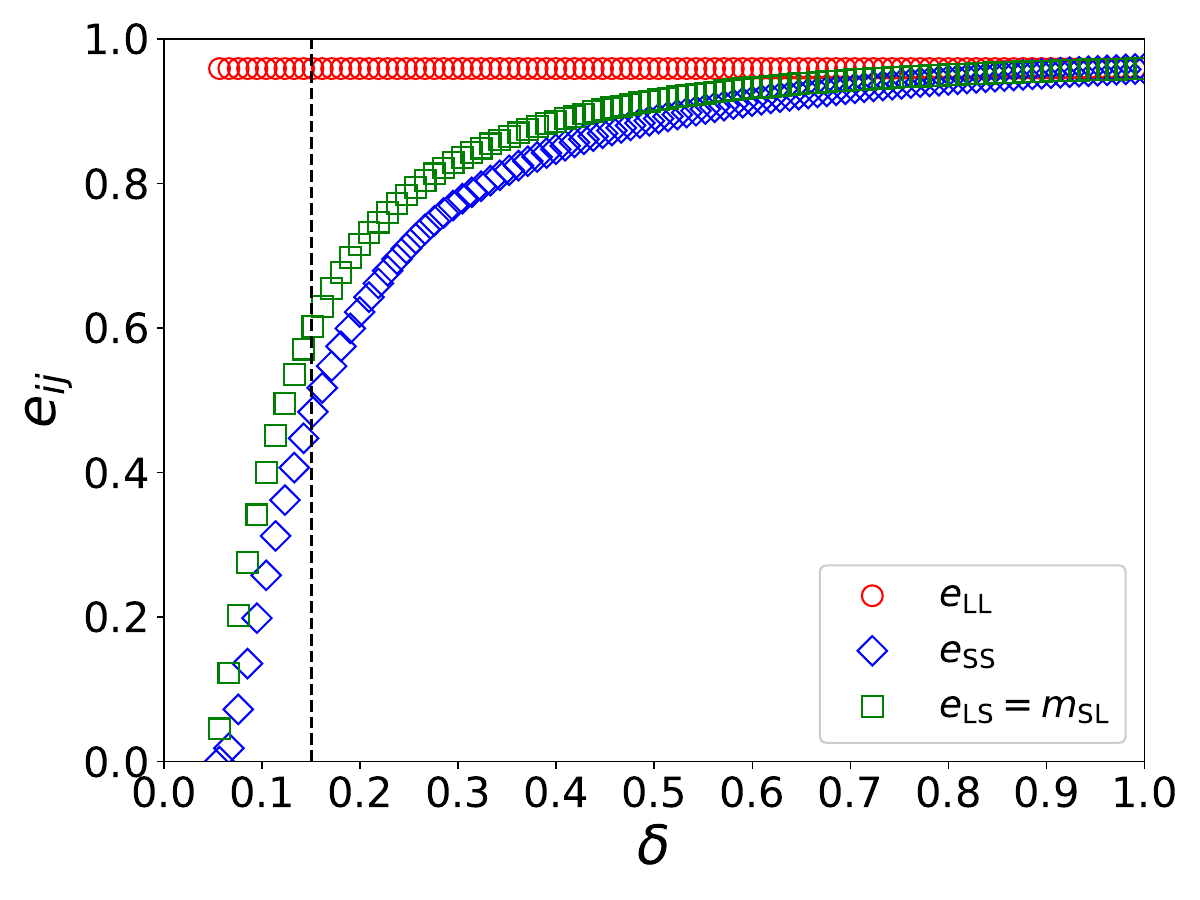}
\caption{Partial restitution coefficients are plotted against the size ratio $\delta$. 
The vertical dashed line corresponds to the smallest value used, $\delta = 0.15$. 
Note that for $\delta \to 1$, 
$e_{\mathrm{SS}}, e_{\mathrm{LS}} \to e_{\mathrm{LL}} = 0.95$. $e_{ij}$  
stops at $\delta = 0.0564$ since below imaginary values are obtained for $e_{\rm SS}$ and 
$e_{\rm LS}$.}
\label{app_fig1}
\end{figure}

The restitution coefficients between particles $i =$ L,S and $j =$ L,S, 
for the linear contact model used, are given by 
\begin{equation}
e_{ij} = \exp \left ( - \dfrac{\gamma_{n}t_{c}^{ij}}{2m_{ij}} \right ) ,
\label{appA1}
\end{equation}
with  the normal damping $\gamma_{n}$, 
the effective mass, $m_{ij}=m_i m_j/(m_i+m_j)$, 
and the contact duration time, $t^{ij}_{c}$, such that:

\begin{align}
m_{\mathrm{LL}} &= \pi/12 = m_{\mathrm{L}}/2 \,,  \label{appA2} \\
m_{\mathrm{SS}} &= m_{\mathrm{LL}}\,\delta^3 \,, \label{appA3}\\
m_{\mathrm{LS}} &= m_{\mathrm{SL}} = 2 m_{\mathrm{LL}} \bigg(\frac{\delta^3}{1+\delta^3}\bigg)\,, \label{appA4}\\
t^{ij}_{c} &= \frac{\pi}{\sqrt{\frac{\kappa_{n}}{m_{ij}} - \big(\frac{\gamma_n}
{2m_{ij}}
\big)^{2}}} ~,
\label{appA5}
\end{align}
where the dimensionless variables $\kappa_{n} = 1$, $\rho_{p} = 1$, 
and $\gamma_{n} = 0.013$ are inserted.

The dependence of $e$ with $\delta$ is shown in Fig.~\ref{app_fig1}. 
We observe that $e_{\mathrm{SS}}$ and $e_{\mathrm{LS}}$ decrease as $\delta \to 0$, 
while $e_{\mathrm{LL}}$ does not change since the radius of the large particles is constant. 
For $\delta = 0.15$ (dashed line), $e_{\mathrm{LL}} = 0.95$,  $e_{\mathrm{SS}} = 0.47$ and  
$e_{\mathrm{LS}} = e_{\mathrm{SL}} = 0.59$, respectively. 
This demonstrates how the dissipation of each contact type depends on $\delta$. 
The fact that large $\delta \rightarrow 1$ results in rather weak dissipation
is one reason for using the background dissipation. However, we believe
that even though competition between the energy dissipated by collisions 
and the energy dissipated by the background medium might take place, 
this should be relevant only in dynamic situations. 
This can be seen in the energy ratio given in Fig.~\ref{protocol} (Bottom), 
where for $\delta = 0.4$ and $\delta = 0.73$, the potential energy dominates over 
kinetic energy, during the whole process, above jamming -- by orders of magnitude.
For static packings created by a slow enough, quasistatic deformation, 
the interplay between dissipated energies should neither play a role in the 
determination of the jamming density \cite{luding2021does} nor for pressure 
and thus bulk modulus.

\section{Partial pressure definitions}
\label{AppB}

The dimensionless pressure and the components of the stress tensor are defined by
\begin{equation}
P = \frac{2r'_{\mathrm{L}}}{\kappa'_{n}} P' = \frac{1}{3}\mathrm{tr}\{\sigma_{\alpha \beta}\} ; \quad \sigma_{\alpha \beta} = \frac{1}{V} \sum_{c \in V} f_{\alpha}\ell_{\beta}.
\label{appB1}
\end{equation}
\noindent where $f_{\alpha}$ is the dimensionless force components and $\ell_{\beta} = r_{\beta}^{j} - r_{\beta}^{k}$ 
are the dimensionless branch vector components connecting the center of the particles $k$ and $j$ that share contact $c$.
The sum over all contacts in the volume $V$ can be decomposed into four sums, each one running 
over each contact type in the system, this means LL, SS, LS, SL. This is written as

\begin{equation}
\begin{split}
\sigma_{\alpha \beta} &= \frac{1}{V} \sum_{i = 1}^{N^{c}_{\mathrm{LL}}} f^{\mathrm{LL}}_{\alpha i}\ell^{\mathrm{LL}}_{\beta i}
					  + \frac{1}{V} \sum_{i = 1}^{N^{c}_{\mathrm{SS}}} f^{\mathrm{SS}}_{\alpha i}\ell^{\mathrm{SS}}_{\beta i}\\
					  &+ \frac{1}{2V} \sum_{i = 1}^{N^{c}_{\mathrm{LS}}} f^{\mathrm{LS}}_{\alpha i}\ell^{\mathrm{LS}}_{\beta i}
					  + \frac{1}{2V} \sum_{i = 1}^{N^{c}_{\mathrm{SL}}} f^{\mathrm{SL}}_{\alpha i}\ell^{\mathrm{SL}}_{\beta i},
\label{appB2}
\end{split}
\end{equation}

\noindent where $N^{c}_{\mathrm{LL}}$, $N^{c}_{\mathrm{SS}}$, $N^{c}_{\mathrm{LS}}$ and $N^{c}_{\mathrm{SL}}$ are the number of 
contacts of each contact type, with $N^{c}_{\mathrm{LS}} = N^{c}_{\mathrm{SL}}$. 
Applying the property $\mathrm{tr}\{A + B\} = \mathrm{tr}\{A\} + \mathrm{tr}\{B\}$, dividing by 3 in 
Eq.(\ref{appB2}) and using the definition of pressure given in Eq.~(\ref{appB1}), one arrives at 

\begin{figure}[t]    
    \centering \includegraphics[scale=0.28]{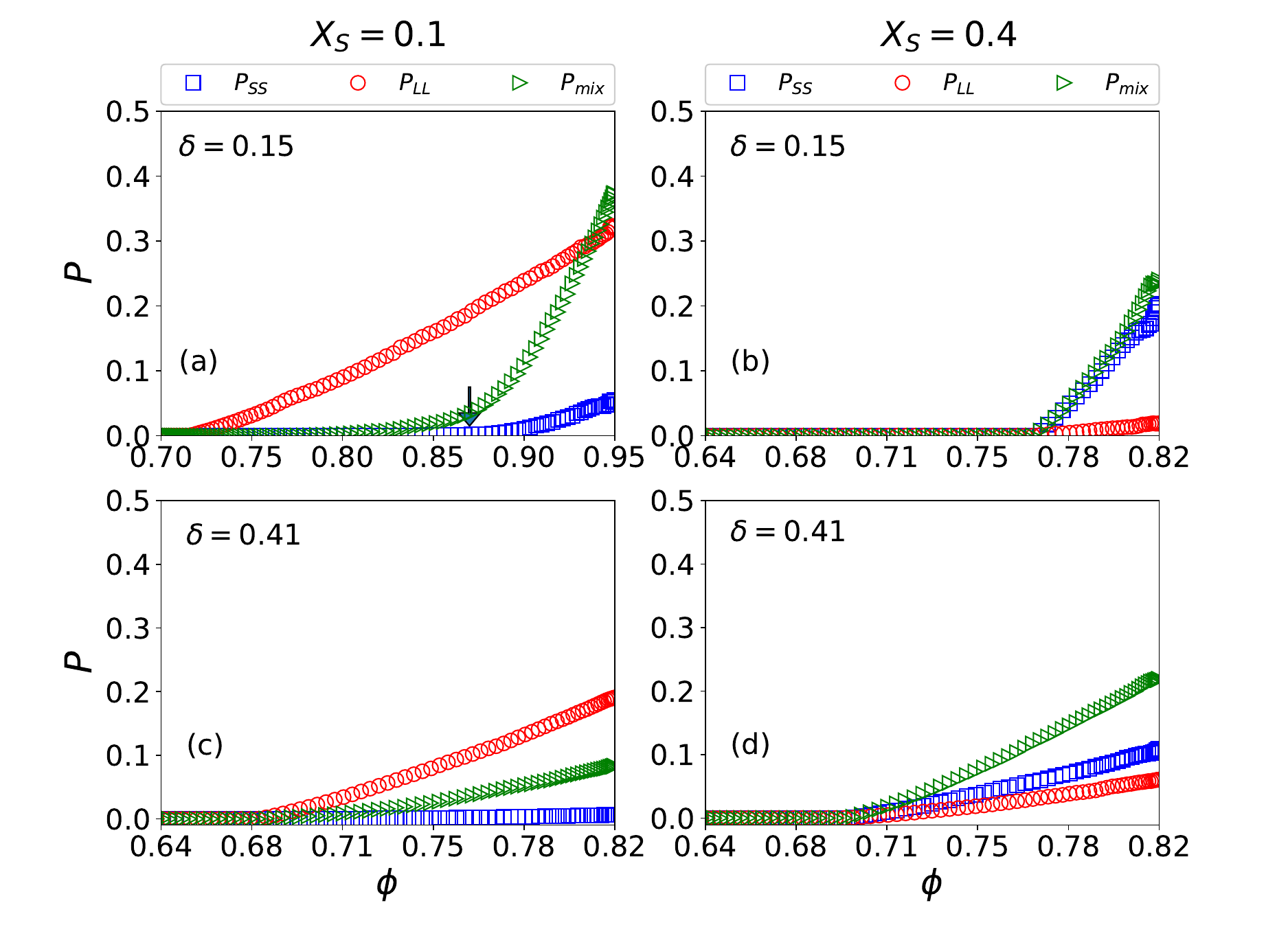}
    \centering \includegraphics[scale=0.28]{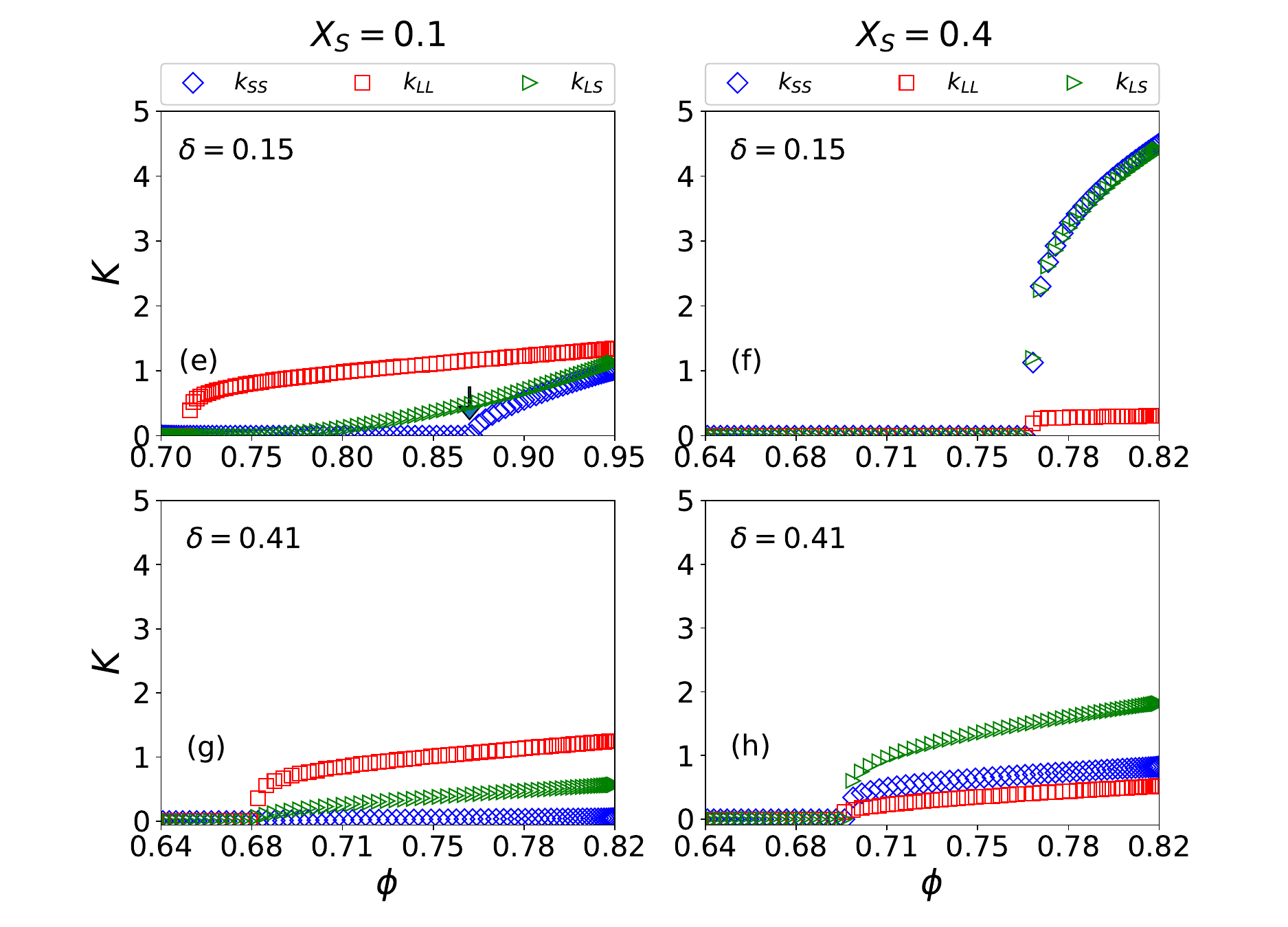}
    \caption{
    (a)-(d) $P_{\mathrm{LL}}$,  $P_{\mathrm{SS}}$ and  $P_{\mathrm{LS}}$ as a function of $\phi$
    for the same $\delta$ and $X_{\mathrm S}$ shown in Fig.~\ref{pressure}. (e)-(h) $K_{\mathrm{LL}}$,  
    $K_{\mathrm{SS}}$ and $K_{\mathrm{LS}}$ as a function of $\phi$ for the same values shown in Fig.~\ref{BulkM}.
	The arrow indicates the second transition, $\phi_J \approx 0.87$, extracted by the derivative of $n_{\mathrm{S}}$, 
	see Fig.~\ref{fractpart_nu}.}
	\label{partialvalues}
\end{figure}
\begin{equation}
P = P_{\mathrm{LL}} + P_{\mathrm{SS}} + \frac{1}{2} (P_{\mathrm{LS}} + P_{\mathrm{SL}}).
\label{appB3}
\end{equation}
where
\begin{align}
P_{\mathrm{LL}} &= \frac{1}{3V} \sum_{i = 1}^{N^{c}_{\mathrm{LL}}} f^{\mathrm{LL}}_{\alpha i}\ell^{\mathrm{LL}}_{\alpha i},
\label{appB4} \\
P_{\mathrm{SS}} &= \frac{1}{3V} \sum_{i = 1}^{N^{c}_{\mathrm{SS}}} f^{\mathrm{SS}}_{\alpha i}\ell^{\mathrm{SS}}_{\alpha i},
\label{appB5}\\
P_{\mathrm{LS}} &= \frac{1}{3V} \sum_{i = 1}^{N^{c}_{\mathrm{LS}}} f^{\mathrm{LS}}_{\alpha i}\ell^{\mathrm{LS}}_{\alpha i},
\label{appB6}\\
P_{\mathrm{SL}} &= \frac{1}{3V} \sum_{i = 1}^{N^{c}_{\mathrm{SL}}} f^{\mathrm{SL}}_{\alpha i}\ell^{\mathrm{SL}}_{\alpha i}.
\label{appB7}
\end{align}

Therefore, $P_{\mathrm{L}} = P_{\mathrm{LL}} + P_{\mathrm{LS}}$ and  
$P_{\mathrm{S}} = P_{\mathrm{SS}} + P_{\mathrm{SL}}$. Since $P_{\mathrm{LS}} = P_{\mathrm{SL}}$, a factor 
of $1/2$ in Eqs.~(\ref{appB2}) and (\ref{appB3}) has to be used 
to avoid an overestimation of the total pressure, 
since the mixed contacts are counted twice. 
$P_{\mathrm{LS}} = P_{\mathrm{SL}}$ is obtained because the normal 
branch vector length is defined between particle centers making no distinction 
between particle sizes, i.e., $\ell^{\mathrm{LS}} = \ell^{\mathrm{SL}}$.
Alternatively, if the branch vector length were defined using the distance 
from the center of each particle $i$ to its contact location, at a distance
radius minus overlap/2 \cite{luding2008cohesive}, then $P_{\mathrm{LS}} \neq P_{\mathrm{SL}}$, 
since $r_{\mathrm{L}} \neq r_{\mathrm{S}}$, 
see Refs.~\cite{fan2011theory,tunuguntla2016discrete}
for the method applied here and the alternative, respectively. 
Discussing the differences between the two stress-definitions
is beyond the scope of this paper and thus postponed to future studies.

Fig.~\ref{partialvalues} (a)-(d) shows the values of the partial pressures calculated using the expressions
given in Eqs.~(\ref{appB4})-(\ref{appB7}). For $\delta = 0.15$ and $X_{\mathrm S} = 0.1$, large particles 
first jam at $\phi \approx 0.71$, then mixed contacts do so (smoothly) at $\phi \approx 0.75$ and later small particles jam at 
$\phi \approx 0.87$. Fig.~\ref{partialvalues} (e)-(h) shows the partial bulk modulus extracted by the fitting method defined in 
Appendix~\ref{AppC}. One can see, that $K_{\mathrm{LL}}$ and $K_{\mathrm{SS}}$ exhibit a jump, whereas 
$K_{\mathrm{LS}}$ increases smoothly, see Fig.~\ref{partialvalues} (e). This demonstrates that the LS-SL 
contact network does not experience a sharp transition whereas the SS contacts do. In fact, $K_{\mathrm{LS}}$ 
dominates the bulk modulus in $K_{\mathrm{S}}$ observed in Fig.~\ref{BulkM} (a), thus hiding
the jump in $K_{\mathrm{SS}}$.

\section{Fitting parameters for the pressure}
\label{AppC}

\begin{figure}[t]
    
    \centering \includegraphics[scale=0.43]{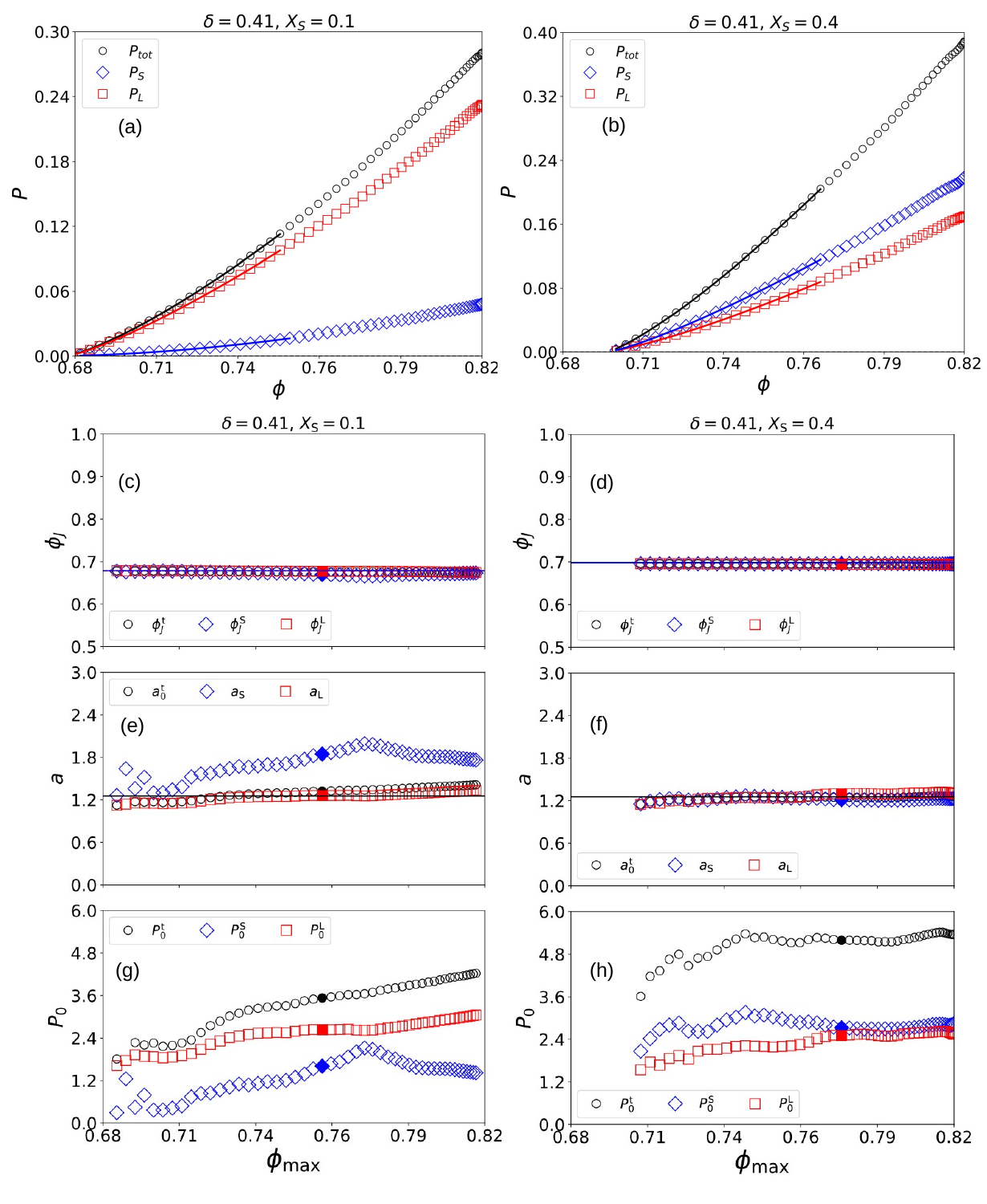}

    \caption{
    Total and partial pressures for size ratio $\delta = 0.41$ with $X_{\mathrm S} = 0.1$ (a) and 
    $X_{\mathrm S} = 0.4$ (b); the corresponding fits are given by $P = P_{0}(\phi - \phi_J)^{a}$, with 
    (c)-(h) fitting parameters ($\phi_J$, $a$, $P_{0}$) as functions of $\phi_{\mathrm{max}}$
    {(where 50 points is the whole regime $\phi \in [\phi_J:\phi_{\rm max}=0.82]$.)} 
    The lines in (c)-(d) represent the $\phi_J$ shown in Fig.~\ref{jamming}.
    The lines in (e)-(f) represent $a_{\mathrm{Mono}} = 1.25$, slightly different from the value $a \sim 1.1$ in Ref.~\cite{majmudar2007jamming}.
    The solid symbols correspond to 
    the values of the parameters obtained by the fits given in (a) and (b) respectively. 
}
     \label{param_SRp4143}
\end{figure}

A power law of the form $P = P_{0}(\phi - \phi_J)^{a}$ is used to fit the 
dimensionless pressure data of each bidisperse packing and study the variation
of the fitting parameters with the fitting range of volume fractions, $\phi$,
to find the range/regime where the parameters are least dependent on the fitting details, 
like the number of points, $n$.
The fitting begins using the first three points of the data, $n = 3$, 
starting around, slightly above $\phi_J$, until a maximum, 
which is varied point by point following 
$\phi_{\mathrm{max}} = \phi_J + n \Delta\phi$, with $\Delta\phi \approx 10^{-3}$.
The fit begins with three points 
since in the least-squares method the number of data points, $n$, must be equal or
higher than the number of fitting parameters, in our case $m = 3$, to find a solution.
For each fit, we extract the values of the fitting parameters as a function of 
$\phi_{\mathrm{max}}$.

\begin{figure}[t]
    
    \centering \includegraphics[scale=0.43]{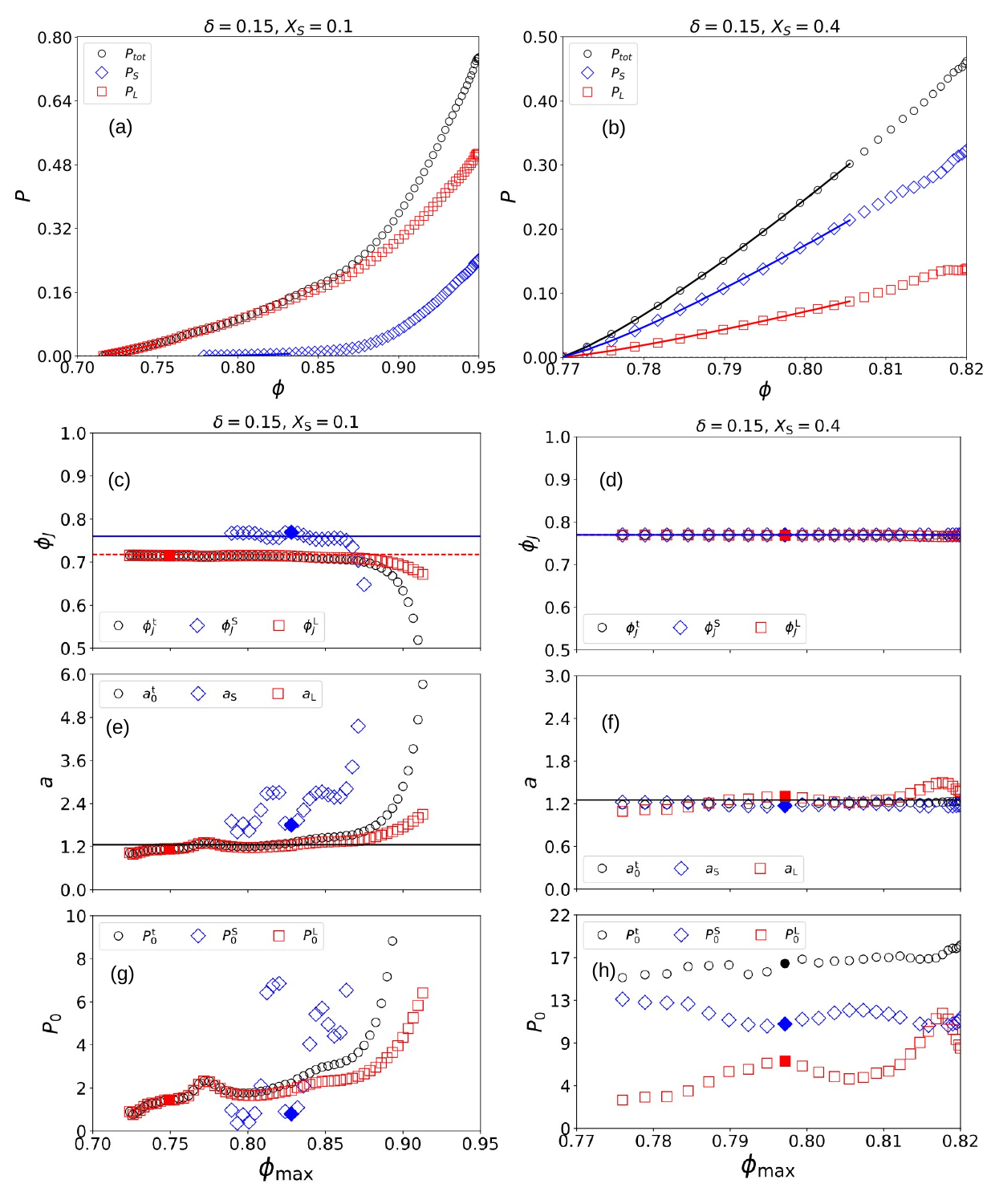}

    \caption{
    Total and partial pressures for size ratio $\delta = 0.15$ with $X_{\mathrm S} = 0.1$ (a) and 
    $X_{\mathrm S} = 0.4$ (b); the other figures and symbols are as 
    in Fig.\ \ref{param_SRp4143}. Note the different number of points, due to the maximum 
    (left, $\phi_{\rm max}=0.95$) and (right, $\phi_{\rm max}=0.82$).
    }
	\label{param_SRp1548}
\end{figure}

Figs.~\ref{param_SRp4143} and \ref{param_SRp1548} show the fits of the
total and partial pressures, and the variation of the fitting parameters $a$, $\phi_J$, 
and $P_0$ as functions of the fit-range, i.e., number of data points, $n$.
For $\delta = 0.41$, the fitting parameters, especially $a$ and $\phi_J$, do not 
change too much, almost consistently for the different species. 
While $a$ does increase a little with $n$, for lower $X_{\mathrm S}$, 
the parameter $a_{\mathrm S}$ turns out to be 
different, i.e., sensible to the fitting range, 
see Fig.~\ref{param_SRp4143}(e). 
The parameter $P_0$
is significantly different between total pressure
and the different species, see 
Fig.~\ref{param_SRp4143}(g) and (h).

For $\delta = 0.15$, $a$ and $\phi_J$ do not show a significant variation for higher $X_{\mathrm S}$. However, 
$a_{\mathrm S}$ is quite sensitive to the fitting range at low $X_{\mathrm S}$, see Fig.~\ref{param_SRp1548}. 
We think that such variations are due to the 
high overlaps developed during compression, 
strongly affecting the power law exponent. 
Most of the fitting parameters of $P_{\mathrm L}$ are rather insensitive to the fitting 
range, thus, the fitting range for $P_{\mathrm L}$ is arbitrary. For $P_{\mathrm S}$ the situation is 
quite different: The parameters are quite sensitive to the fitting range at low $\delta$ 
and low $X_{\mathrm S}$. Therefore, a special assumption is considered for the fitting
range, i.e., we fit the data of $P_{\mathrm S}$, at low $\delta$ and low
$X_{\mathrm S}$, with a narrow fitting range near to $\phi_J$ keeping $a_{\mathrm S} \sim a_{\mathrm L}$, as closely
as possible. This guarantees low overlaps for small particles and similar power law exponents for both large and small particles. 
The fitting parameters extracted from $P_{\mathrm L}$ and $P_{\mathrm S}$ are used to determine
the bulk modulus of large and small particles, using $K = \phi P_{0}a(\phi - \phi_{J})^{a-1}$, 
in Sec.~\ref{SecIV}.

\section{Linear relation between pressure $P$ and overlap $\langle \alpha \rangle$}
\label{AppD}

\begin{figure}[h]
    \centering \includegraphics[scale=0.27]{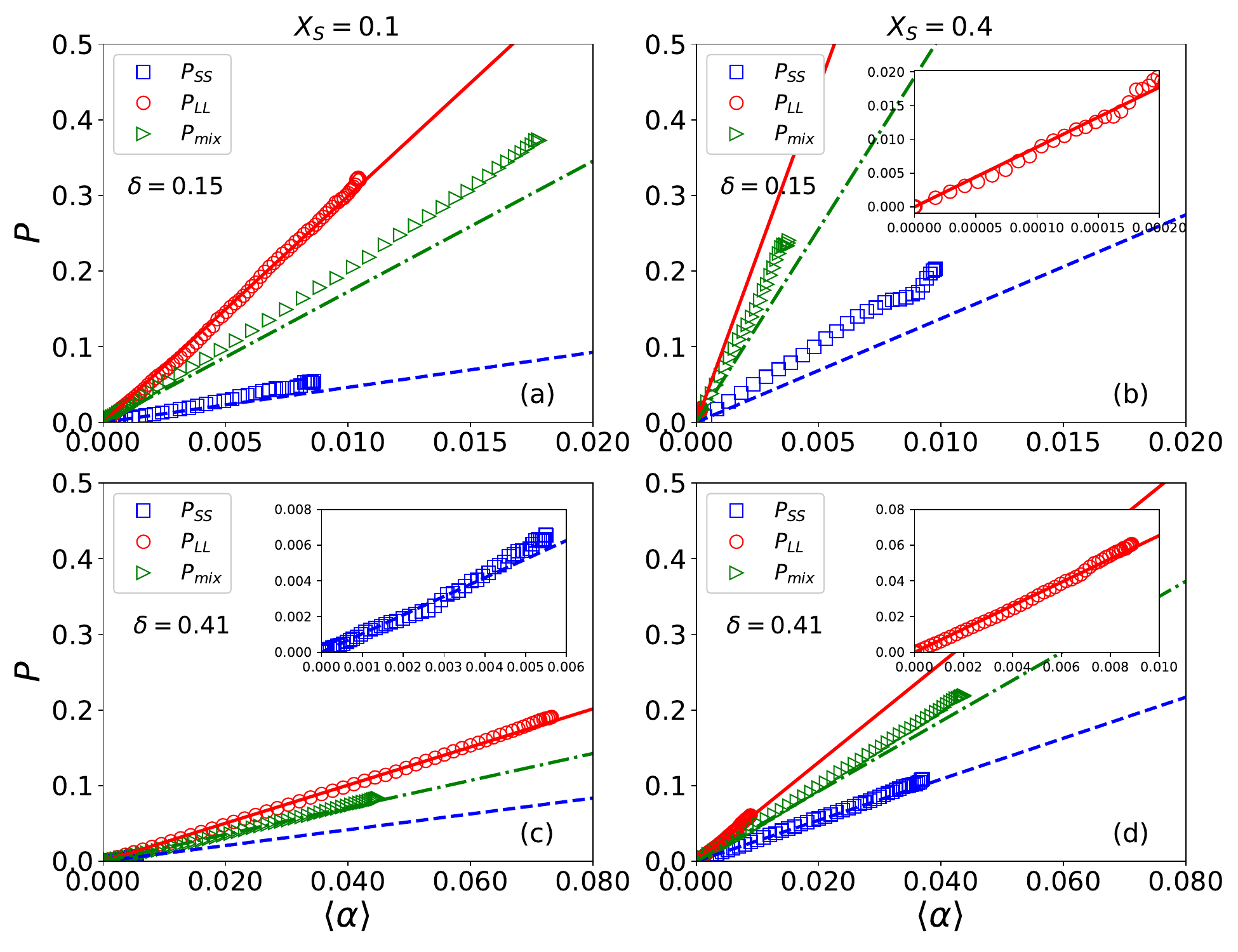}
    \caption{The partial dimensionless pressures $P=P_{ij}$ as functions of the dimensionless mean 
    overlap $\langle \alpha \rangle = \langle \alpha_{ij} \rangle$ for different $\delta$ and $X_{\mathrm S}$ values. Note the different horizontal axis ranges for different $\delta$.
    }
	\label{p_vs_alpha}
\end{figure}

In this section, we demonstrate that the dimensionless partial pressures, $P_{ij} = 2 r^{\prime}_{\rm L} P^{\prime}_{ij}/\kappa'_{n}$, 
depend linearly on
the dimensionless mean overlap,
$\langle \alpha_{ij} \rangle = \langle \alpha_{ij}^{\prime} \rangle /2 r^{\prime}_{L}$, 
where $i,j \in$ [L,S]. This is shown in Fig.~\ref{p_vs_alpha} for the same values of $\delta$ and 
$X_{\mathrm{S}}$ as in Fig.~\ref{partialvalues} (a)-(d).

The ranges of pressures and mean overlaps depend on the contact types. Each contact type experiences different overlaps at different pressures, evidencing the different roles each 
contact type has inside the force network. 
To understand why pressure is linear with overlap, one can write 
$P_{\mathrm{LL}}
= 2 P_{\mathrm{LL}}'r_{\mathrm{L}}'/\kappa_n'
\approx (Z_{\mathrm{LL}}/3V')\, 4 F_{\mathrm{LL}}'r^{\prime 2}_{\mathrm{L}}/\kappa_n' 
= (4/3) Z_{\mathrm{LL}} \langle \alpha'_{\mathrm{LL}}\rangle/r_{\mathrm{L}}' 
\propto Z_{\mathrm{LL}} \langle \alpha_{\mathrm{LL}}\rangle$, using $V'\approx r_{\mathrm{L}}^{3\prime}$ and $F_{ij}'=\kappa_n' \langle \alpha'_{ij} \rangle$. Using similar reasoning, we find 
$P_{\mathrm{SS}} 
\propto Z_{\mathrm{SS}} \langle \alpha_{\mathrm{SS}}\rangle \delta$
and 
$P_{\mathrm{LS}} 
\propto Z_{\mathrm{LS}} \langle \alpha_{\mathrm{LS}}\rangle (1+\delta)/2$ due to the branch vectors $\ell'_{ij}=r'_i+r'_j$. The trend lines
in Fig.~\ref{p_vs_alpha} quantify 
the different slopes that are due to the different partial mean contact numbers $Z$,
the different mean overlaps, and the different branch vectors. For $X_{\mathrm{S}}=0.1$, the large and the mix contribution is dominating, the small particles contribute little for $\delta=0.41$ and only a bit for $\delta=0.15$. In contrast, for $X_{\mathrm{S}}=0.4$, the small and the mix contribution is dominating, the large particles hardly contribute for $\delta=0.15$ and only little for $\delta=0.41$.

\bibliography{Ref,new0122}
\bibliographystyle{apsrev4-1}

\end{document}